\documentclass[aps,prb,twocolumn,amsmath,amssymb,superscriptaddress,floatfix,eqsecnum,showpacs,nofootinbib]{revtex4}
\usepackage{amsmath}
\usepackage{graphicx}
\usepackage{latexsym}
\usepackage{amsmath,amssymb}
\usepackage{bm} 
\usepackage{color}
\usepackage{stackrel}
\DeclareMathOperator{\sgn}{sgn}

\newcommand{\bsub}{\begin{subequations}}
\newcommand{\esub}{\end{subequations}}

\newcommand{\st}[1]{\scriptstyle{#1}}
\newcommand{\sss}[1]{\scriptscriptstyle{#1}}
\newcommand{\msf}[1]{\mathsf{#1}}

\newcommand{\ord}[1]{\bm{\mathit{O}}\left(#1\right)}
\newcommand{\sqord}[1]{\bm{\mathit{O}}\left[#1\right]}
\newcommand{\intl}[1]{\int\limits_{#1}}
\newcommand{\vex}[1]{\bm{\mathrm{#1}}}

\newcommand{\sigh}{\hat{\sigma}}
\newcommand{\sigb}{\hat{\bm\sigma}}

\newcommand{\T}{\mathsf{T}}

\newcommand{\ket}[1]{\left| {#1} \right\rangle}

\newcommand{\loc}{\xi_{\msf{loc}}}

\newcommand{\vf}{v_{F}}
\newcommand{\dmo}{\Delta_M^{{\sss{(}}\circ{\sss{)}}}}
\newcommand{\dvo}{\Delta_{V}^{{\sss{(}}\circ{\sss{)}}}}
\newcommand{\dAo}{\Delta_A^{{\sss{(}}\circ{\sss{)}}}}
\newcommand{\parl}{{\st{\parallel}}}
\newcommand{\nimp}{n_{\msf{imp}}}

\newcommand{\typ}[1]{{#1}_{\mathsf{typ}}}

\newcommand{\tnu}{\tilde{\nu}}

\newcommand{\numax}{\nu_{\mathsf{max}}}
\newcommand{\numin}{\nu_{\mathsf{min}}}

\begin{document}

\title{
Multifractal nature of the surface local density of states in three-dimensional topological insulators
with magnetic and nonmagnetic disorder
}

\author{Matthew S. Foster}
\email{psiborf@rci.rutgers.edu}
\affiliation{Center for Materials Theory, Department of Physics and Astronomy,
Rutgers University, Piscataway, NJ 08854, USA}

\date{\today}

\begin{abstract}
We compute the multifractal spectra associated to local density of states (LDOS) fluctuations due to weak 
quenched disorder, for a single Dirac fermion in two spatial dimensions. Our results are relevant
to the surfaces of $\mathbb{Z}_2$ topological insulators such as $\text{Bi}_2\text{Se}_3$ and $\text{Bi}_2\text{Te}_3$, 
where LDOS modulations can be directly probed via scanning tunneling microscopy. We find a 
qualitative difference in spectra obtained for magnetic versus non-magnetic disorder. Randomly polarized 
magnetic impurities induce quadratic multifractality at first order in the impurity density; by 
contrast, no operator exhibits multifractal scaling at this order for a non-magnetic impurity profile. 
For the time-reversal invariant case, we compute the first non-trivial multifractal correction, which appears 
at two loops (impurity density squared). We discuss spectral enhancement approaching the Dirac point due to 
renormalization, and we survey known results for the opposite limit of strong disorder.
\end{abstract}


\maketitle


\section{Introduction}

The defining attribute of 
a
3D $\mathbb{Z}_2$ topological insulator\cite{3DTopIns} (TI) is the
presence of an odd number of 2D massless Dirac bands at the material surface.\cite{KaneHasanREVIEW,ZhangREVIEW}
Unlike the Dirac electrons that can appear in a purely 2D system (notably in graphene),
the surface states of a (strong) 3D TI are robustly protected from the
opening of gap, so long as time-reversal symmetry is preserved. The protection
can be viewed as a consequence of the parity anomaly,\cite{Semenoff,RedlichJackiw,Haldane88,ZhangREVIEW} 
which ``holographically'' links surface states separated by a topologically non-trivial bulk, 
and gives rise to the signature properties of the $\mathbb{Z}_2$ TI 
state:
the half-integer quantum Hall effect, 
quantized magnetoelectric coupling, ``axion'' electrodynamics, etc.\cite{KaneHasanREVIEW,ZhangREVIEW}
As stressed by Schnyder et al.\ in Ref.~\onlinecite{SRAL08}, the robust character of 
the surface states in the presence of quenched disorder can also be taken 
as a principal characteristic
of a topological insulator. In particular, these states are protected from
Anderson localization,\cite{LeeRamakrishnan} 
even in the presence of 
a ``strong'' impurity potential,
so long as time-reversal invariance
is preserved.\cite{Z2Shinsei,Z2Numerics}

\begin{figure}[b!]
   \includegraphics[width=0.41\textwidth]{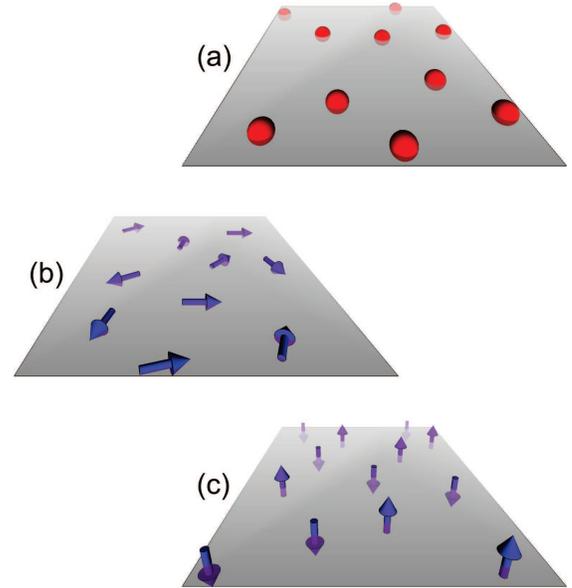}
   \caption{
	Sketch of disorder ``flavors'' on the surface of a 
	$\mathbb{Z}_2$ topological insulator.
	In the time-reversal invariant case, the impurities
	are neutral adatoms or charged dopant ions, depicted as
	spheres in {\bf (a)}. The effects of these on the surface
	Dirac theory [Eq.~(\ref{H})] are encoded in the scalar 
	potential $V(\vex{r})$. In the case of magnetic disorder, 
	the impurity spins are indicated by the arrows in
	{\bf (b)} and {\bf (c)}. In the limit that the spins reside in 
	the plane of the surface, {\bf (b)}, the disorder appears as a vector
	potential $\vex{A}(\vex{r})$. The opposite case of
	out-of-plane polarization, {\bf (c)}, gives the 
	random mass $M(\vex{r})$. The case of generic time-reversal 
	breaking disorder has all three potentials present.
	}
   \label{DirtVar}
\end{figure}

With its 2D Dirac band pinned to an exposed surface, a 3D TI is ideally 
suited to local probes such as scanning tunneling microscopy (STM). 
In spectroscopic mode, an STM captures an areal map of the local density of states (LDOS). 
There are several ways of analyzing such data. One is to look for quasiparticle 
interference (QPI)\cite{Capriotti03,GuoFranz10,QPI-TI-SingImp,QPI-STM,Yazdani11} 
in the LDOS Fourier transform. This method is useful for determining short-distance 
details, and contains 
similar
information as an analysis of LDOS Friedel oscillations in the presence of a 
single impurity.\cite{SingImpRef} It has been applied in TIs to experimental data 
and analyzed theoretically in Refs.~\onlinecite{QPI-STM,Yazdani11} and \onlinecite{GuoFranz10,QPI-TI-SingImp}, 
respectively. In QPI, the disorder is employed primarily as a facilitator to gleam 
information about the \emph{clean} system.\cite{Capriotti03}

Multifractal analysis\cite{MFCReview,Mirlin,PaladinVulpiani} provides a complementary method better
suited to extracting large-distance, disorder-dominated features
in the same LDOS data field. 
It is a standard tool for assaying quantum interference phenomena, and is employed
in the analysis of wavefunctions near a metal-insulator transition\cite{Wegner,Pruisken,MFCReview,Mirlin}
as well as mesoscopic fluctuations in diffusive metallic systems.\cite{Boris,FalkoEfetov95} 
In this paper, we derive new results for LDOS multifractal spectra associated to
disordered topological insulator surface states. In particular, we extend the pioneering
results of Ref.~\onlinecite{Ludwig94} to the generic cases of time-reversal 
($\mathcal{T}$)
preserving and breaking impurities. Our calculations are performed in the near-ballistic limit,\cite{Schuessler09}
wherein weak disorder enters as a perturbation to the clean Dirac band structure. 
A key characteristic of 2D Dirac fermions is that this weak disorder regime is continuously 
connected to more conventional domains of multifractal analysis, i.e.\ the diffusive (symplectic)
metal\cite{Wegner,FalkoEfetov95} and the integer quantum Hall plateau 
transition.\cite{MFCReview,PookJanssen91,KlesseEvers95,ObuseEvers08} 
These appear at \emph{strong} coupling (many impurities) for dirty Dirac fermions.\cite{Ludwig94,Z2Shinsei,Z2Numerics,NomuraRyu08}

We consider
the case of a single flavor Dirac surface band, 
relevant to (e.g.) the TIs Bi${_2}$Se$_{3}$ and Bi${_2}$Te$_{3}$ 
(Refs.~\onlinecite{KaneHasanREVIEW,ZhangREVIEW,Bi2X3 BS}).
The different kinds of $\mathcal{T}$-preserving and $\mathcal{T}$-breaking
disorder are sketched in Fig.~\ref{DirtVar}.
We
demonstrate that the LDOS multifractal spectra observed in the absence of time-reversal 
symmetry breaking (i.e., for non-magnetic disorder) is qualitatively weaker than that
induced by magnetic impurities. 
In particular, the first multifractal correction
obtains at first order in the impurity density for the case of broken $\mathcal{T}$,
while the first non-trivial 
amplitude
appears at second order in the $\mathcal{T}$-invariant case. We compute the 
leading 
terms
via one- and two-loop calculations, respectively. We also compute unnormalized
spectra for the spin LDOS\cite{spinLDOS} in the case of magnetic impurities.
We 
show that renormalization effects can enhance
multifractality near the Dirac point. 
Finally,
we summarize prior results on various strong-coupling regimes.
Our goal is to
sketch the full portrait of quantum interference physics on the surface of a TI, 
valid when interparticle interactions can be neglected.

Our results indicate that the long-distance, disorder-dominated features
captured by the multifractal analysis behave in many cases opposite to the short-distance
characteristics that appear in quasiparticle interference.\cite{GuoFranz10,QPI-TI-SingImp,QPI-STM,Yazdani11} 
In Ref.~\onlinecite{GuoFranz10}, the authors observed that QPI
is strongest for the spin LDOS response to magnetic impurities, while the
unpolarized LDOS pattern \emph{vanishes} for magnetic disorder (in the first Born approximation).
The QPI response of the LDOS to non-magnetic disorder is weak but non-zero.\cite{GuoFranz10}
By contrast, in this work we find that the LDOS 
multifractality
is 
strongest for magnetic impurities, while the spin LDOS spectrum comparatively 
exhibits the same or weaker strength fluctuations, depending upon the polarization direction.

The weak influence of non-magnetic disorder is tied to the intrinsic spin-orbit coupling
that defines the massless Dirac kinetic term. Multifractality is suppressed at one loop due 
to interference mediated by the Dirac pseudospin, which is proportional to the physical spin
on a $\mathbb{Z}_2$ insulator surface.
The spin is also responsible for the suppression of backscattering from a single non-magnetic impurity.\cite{SuppBackScat}
On the TI surface, magnetic disorder Zeeman couples directly to the Dirac spin, 
enabling backscattering in near-ballistic transport, and inducing multifractal LDOS fluctuations 
at the lowest order in the impurity density.

A notable problem in experiments probing topological insulator surface states has been
the unintentional doping of carriers into the bulk bands, which then dominate
transport measurements in large 
samples.\cite{Transport}
Even if the chemical potential is moved into the gap, it may reside far 
from the Dirac point, making it difficult to observe surface state carrier 
dynamics at low densities. In this respect, STM offers several advantages 
over transport experiments. First, the position of the chemical potential 
is no barrier to probing states at the Dirac point, since the latter can always be reached
by tuning the bias voltage (although the Dirac point is not guaranteed
to reside in the bulk gap).\cite{ZhangREVIEW,Bi2X3 BS} 
Assuming 
that
the Dirac point
or the 
low density regime can 
be accessed
by tuning the tunneling bias,
the advent of a finite, even large doping of the 
surface and/or bulk states 
may actually play a beneficial
role in facilitating the observation of disorder-induced quantum interference 
effects. This is because a finite carrier density screens the long-range Coulomb 
potential introduced by charged defects.
The
potential landscape formed by 
screened
impurities 
is
short-range correlated
on scales larger than the screening length.
Good screening eliminates 
the problem of electron and hole puddle 
formation,\cite{Puddles,PuddlesReview} 
which has 
until recently\cite{PuddlesNoMore}
occluded transport and other properties of Dirac carriers in graphene near the 
Dirac point.
On the other hand, a low density of poorly-screened bulk dopants
induces a long-range correlated potential and puddle formation, as 
in graphene.\cite{Yazdani11} LDOS fluctuations
in the puddle regime are an important topic for future work.

Three-dimensional topological 
insulators provide us with an interesting paradigm 
flip for quantum interference phenomena.
Isolating the surface state contribution in transport 
measurements is problematic.
By comparison, direct LDOS imaging is easier than in conventional
semiconductor systems, wherein the 2D electron
gas is typically buried in a layered material stack.
Moreover, the amount of surface disorder can to some
extent be controlled; for example, magnetic impurities can be deposited
across the surface of an otherwise high-quality bulk 3D sample.
These can be charge-neutral adatoms or charged dopants; an example of the
former (latter) is provided by iron (manganese)\cite{Shen10} 
in $\text{Bi}_2\text{Se}_3$.

This paper is organized as follows.
We begin in Sec.~\ref{Sec: MFC Meas} with a lightning review of multifractal
composite and spin LDOS measures. In Sec.~\ref{Sec: Weak}, we present
new results for multifractal LDOS fluctuations in TI surface states,
in the presence of weak disorder. We also show how renormalization can 
enhance multifractality close to the Dirac point.
Finally, in
Sec.~\ref{Sec: Strong}, we review previous results on various 
strong disorder regimes relevant to the $\mathbb{Z}_2$ TI
surface states and LDOS statistics. In particular, we discuss the
symplectic metal, the integer quantum Hall plateau transition, and the
Anderson insulator.
Various technical details are relegated to appendices. 
In Appendix~\ref{App--Sym}, we review the symmetry classes of Anderson 
(de)localization that appear in the disordered Dirac surface theory.
In Appendix~\ref{App--PT}, we supply
some details of our perturbative calculations.

\section{Multifractal LDOS measures \label{Sec: MFC Meas}}

\subsection{Definitions}

We suppose that the tunneling local density of states (LDOS) 
$\nu(\varepsilon,\vex{r})$ is imaged at a fixed energy $\varepsilon$ 
over an $L \times L$ field of view. The field is finely partitioned into 
a grid of boxes. The box edge length $a \ll L$ 
must
be chosen larger than 
any ``microscopic'' scale $l_{\msf{m}}$, such as the correlation 
length of the random potential.\cite{MFCReview} One introduces the box probability
\begin{align}\label{Box Prob}
	\mu_n(\varepsilon) 
	\equiv
	\frac{
	\intl{\mathcal{A}_n}
	d^{2}\vex{r} \,
	\nu(\varepsilon,\vex{r})
	}
	{
	\sum_{l} 
	\left[
	\intl{\mathcal{A}_l}
	d^{2}\vex{r} \,
	\nu(\varepsilon,\vex{r})	
	\right]
	},
\end{align}
where $\mathcal{A}_n$ denotes the $n^{\text{th}}$ box.
LDOS multifractality is defined through 
the inverse of the participation ratio (IPR),\cite{Wegner,MFCReview}
\begin{align}\label{IPR}
	\mathcal{P}_{q}(\varepsilon)
	\equiv
	\sum_{n} \mu_n^q(\varepsilon)
	\sim
	\left(
	\frac{a}{L}
	\right)^{\tau(q,\varepsilon)}.
\end{align}
The right-hand side (scaling limit) obtains when 
$l_{\msf{m}} \ll a \ll L$; corrections are down by
higher powers of $a/L$. The exponent $\tau(q,\varepsilon)$ 
is the multifractal moment spectrum\cite{Halsey,MFCReview}
for LDOS fluctuations at energy $\varepsilon$.

The construction in Eqs.~(\ref{Box Prob}) and (\ref{IPR}) is useful for characterizing 
a system with extended states,
or for an Anderson localized 
system in which $L \ll \loc(\varepsilon)$; $\loc$ denotes the localization length. In what follows, 
we assume experiments are performed at sufficiently low temperatures so that inelastic cutoffs to 
quantum interference can be ignored.\cite{LeeRamakrishnan,WRM} A clean system with plane wave 
states at energy $\varepsilon$ has $\tau(q,\varepsilon) = 2(q - 1)$. \emph{Multifractality} refers 
to the incorporation of corrections non-linear in $q$. Physically, these 
arise due to quantum interference via multiple scattering of electron waves in a dirty environment, 
processes that serve as the precursor to Anderson localization.\cite{Wegner,MFCReview,Boris}

For weak disorder, the spectrum is typically dominated by the quadratic correction\cite{Wegner,MFCReview}
\begin{align}\label{tauQuad}
	\tau(q,\varepsilon) = 2 (q - 1) - \theta(\varepsilon) \, q(q - 1),
\end{align} 
where $\theta \geq 0$ gives a measure of the disorder strength. 
As an example, in a weakly disordered 2D metal (with $L \ll \loc$ for the orthogonal or unitary classes),
one finds\cite{Wegner,Pruisken,FalkoEfetov95}
\begin{align}\label{tauMetal}
	\theta = \frac{\beta^{-1}}{2 \pi^2 N(\varepsilon) D},
\end{align}
where $N(\varepsilon)$ denotes the average density of states, $D$ is the classical (Drude)
diffusion constant, and 
$\beta \in \{1,2,4\}$, 
depending upon the presence or absence 
of time-reversal symmetry and spin-orbit scattering.\cite{FalkoEfetov95,RandomMatrix}
At stronger disorder, higher order corrections in $\theta q$ must be retained;
for the diffusive metals, results are known to four loops.\cite{Wegner} 

An alternative characterization of LDOS multifractality is provided by the 
singularity spectrum\cite{Halsey,MFCReview} $f(\alpha)$: 
Over a subset of the sample grid
area that scales as $(L/a)^{f(\alpha)}$, the box probability $\mu \sim (a/L)^{\alpha}$.
The singularity spectrum is the Legendre transform of $\tau(q)$,
\[
	f(\alpha) = q \alpha - \tau(q), \quad \frac{d \tau(q)}{d q} = \alpha.
\]
For the quadratic spectrum in Eq.~(\ref{tauQuad}), one obtains
\begin{align}\label{fPara}
	f(\alpha) = 2 - \frac{1}{4 \theta}\left(\alpha - 2 - \theta\right)^2.
\end{align}
In this ``parabolic approximation,'' the strength of the multifractality is encoded
in the peak position $\alpha_0$ [$f(\alpha_0) = 2$], and the width $\alpha_\msf{W}$ 
of the spectrum such that $f(\alpha_0 \pm \alpha_\msf{W}/2) = 0$,
\begin{align}\label{ShiftWidth}
	\alpha_0 = 2 + \theta,\quad \alpha_{\msf{W}} = 4 \sqrt{2 \theta}.
\end{align}

Part of the power of multifractal analysis for disordered
quantum systems derives from the fact that the spectra [$\tau(q)$ or $f(\alpha)$]
typically depend only upon a few gross measures of the impurity potential. In the case
of dirty metals, the entire spectrum can be computed as an expansion in one parameter,
the inverse conductance (consistent with scaling theory).\cite{Wegner,LeeRamakrishnan,Boris} 
At a non-interacting Anderson localization transition, $\tau(q)$ and $f(\alpha)$ become 
universal functions, so that the critical point is characterized by an \emph{infinite} set 
of critical exponents [e.g., the expansion coefficients for $\tau(q)$].

The spectra above have been defined for data collected in a single fixed realization
of the disorder. Strictly speaking, Eq.~(\ref{tauQuad}) then applies only for $|q| \leq q_c$,
where $q_c = \sqrt{2/\theta}$. Outside of this range, the $\tau(q)$ associated to
a fixed disorder realization is linear, a phenomenon known as spectral termination.\cite{Chamon96,Castillo97,Foster09}
[This assumes that higher order corrections can be ignored for $q \geq q_c$. Regardless, the
$\tau(q)$ spectrum is always linear for sufficiently large $q$].
Termination can be viewed as a consequence of the restriction to positive sample measures $f(\alpha) \geq 0$.\cite{Halsey,Mirlin}

In the localized regime, the states contributing to the LDOS at a given position in the sample
have a discrete energy spectrum, quantized by the typical localization volume $\loc^2$. 
As a result, all non-unity LDOS moments \emph{diverge} in the absence of level smearing.
In a tunneling experiment, smearing can appear due to inelastic scattering (temperature), 
open sample boundary conditions, or due to the finite energy resolution of the instrument. 
To characterize an Anderson insulating state over an $L \times L$ field of view with $L \gg \loc$, the full LDOS 
distribution should be examined;\cite{AltPrig87,Lerner88} sensitive dependence of the distribution shape to smearing 
can serve as a telltale sign of the localized regime. LDOS fluctuations in the Anderson insulator
are reviewed in more detail in Sec.~\ref{Sec: Anderson Ins}.

\subsection{Spin LDOS spectra}

By restricting the character of the tunneling species,
it may be possible to measure individual LDOS components separately.
For example, in the case of a spin-polarized (ferromagnetic) STM 
tip,\cite{spinLDOS}
the spin-projected components $\nu_{\uparrow,\downarrow}$ can be separately resolved. 
The use of an unpolarized tip recovers the composite LDOS 
$\nu = \nu_{\uparrow} + \nu_{\downarrow}$.
We define the spin LDOS along the spin space direction $\hat{\iota}$,
\begin{align}\label{SpinLDOS}
	\nu^{\hat{\iota}}(\varepsilon,\vex{r})
	\equiv
	\nu^{\hat{\iota}}_{\uparrow}(\varepsilon,\vex{r})
	-
	\nu^{\hat{\iota}}_{\downarrow}(\varepsilon,\vex{r}).
\end{align}
For a time-reversal invariant system (with or without spin-orbit scattering 
and/or disorder), one has $\nu^{\hat{\iota}}(\varepsilon,\vex{r}) = 0$.
In a system with broken time-reversal (e.g., magnetic impurities), but
zero average spin polarization, the integral of $\nu^{\hat{\iota}}(\varepsilon,\vex{r})$
over a sufficiently large region becomes arbitrarily small; 
we cannot use the normalized construction in Eqs.~(\ref{Box Prob}) and (\ref{IPR}) 
to characterize spin LDOS multifractals. Instead, we employ the un-normalized 
inverse spin participation ratio (ISPR)
\begin{align}\label{ISPR}
	\mathcal{P}^{\hat{\iota}}_{q}(\varepsilon)
	\equiv
	\sum_{n}
	\left(\mu_{n}^{\hat{\iota}}\right)^q,
	\quad
	\mu_{n}^{\hat{\iota}}
	\equiv
	\intl{\mathcal{A}_n}
	d^{2}\vex{r} \,
	\nu^{\hat{\iota}}(\varepsilon,\vex{r}).
\end{align}
In the scaling limit,
\begin{align}\label{ISPR--Scaling}
	\mathcal{P}^{\hat{\iota}}_{q}(\varepsilon)
	\sim
	c_q
	\left(
	\frac{a}{L}
	\right)^{x_{q}^{\hat{\iota}} - 2},
\end{align}
where the exponent $x_{q}^{\hat{\iota}}$ is the scaling
dimension for the corresponding moment operator in the
disorder-averaged field theory description, and 
$c_q \neq 0$ for even $q$.

\section{Weak disorder multifractality \label{Sec: Weak}}

\subsection{Model. Short- and long-range correlated potential landscapes\label{Model}}

The Dirac surface states of a $\mathbb{Z}_2$ topological insulator (TI) are guaranteed
to appear in an odd number of flavors.\cite{KaneHasanREVIEW,ZhangREVIEW} 
In this paper, we consider the simplest case of a single flavor, relevant to (e.g.) 
$\text{Bi}_2\text{Se}_3$ and $\text{Bi}_2\text{Te}_3$. The Hamiltonian is
(in units such that $\hbar = 1$)
\begin{align}\label{H}
	H 
	= 
	\!
	\int \! d^2\vex{r}
	\,
	\psi^\dagger
	\!
	\left\{
	\begin{aligned}
	& \vf \,
	\sigh_{\mu}
	\left[
	-i 
	\partial_{\mu}
	+
	A_\mu(\vex{r})
	\right]
	\\&
	\,+
	M(\vex{r}) \,
	\sigh_3
	+
	V(\vex{r})
	\end{aligned}
	\right\}
	\!
	\psi,
\end{align}
where $\mu \in \{1,2\}$, and repeated indices are summed. 
The coordinates $\vex{r} = \{x,y\}$ chart the TI surface, while the topological bulk
resides in the perpendicular $z$ direction. 
In Eq.~(\ref{H}), $\vf$ denotes the Fermi velocity, and the Dirac pseudospin Pauli matrices $\sigb$ are related
to the physical spin $1/2$ operators $\hat{\bm{S}}$ via 
$\{\sigh_{\mu},\sigh_3\} = 2 \{\epsilon_{\mu \nu} S_{\nu},S_3\}$.
The vector, scalar, and mass potentials $\{\vex{A},V,M\}$ describe the effects of external 
electromagnetic fields and/or surface impurities. 
In the absence of time-reversal ($\mathcal{T}$) symmetry breaking, $\vex{A} = M = 0$.
(See Appendix~\ref{App--Sym} for an enumeration of discrete symmetry operations.)
Thus, a mass gap is explicitly forbidden so long as $\mathcal{T}$ remains a good symmetry,
a consequence of the protection afforded by the topologically nontrivial bulk. 
When $\mathcal{T}$ is broken by an external magnetic field $\vex{B}$, the vector and mass potentials are 
\begin{align}\label{VectorMass}
	A_{\mu} 
	&= 
	-e A_{\mu}^{(\msf{orb})}
	-
	\frac{\gamma_{\parl}}{2 \vf} 
	\epsilon_{\mu \nu} 
	B_{\parl, \nu},
	\nonumber\\
	M
	&=
	-\frac{\gamma_{\perp}}{2}
	B_z,
\end{align}
where $\gamma_{\parl}$ ($\gamma_{\perp}$) denotes the Zeeman coupling to the in-plane field
$\vex{B}_{\parl}$ (out-of-plane field $B_z$), and the orbital effect
is embedded in $A_{\alpha}^{(\msf{orb})}$ via $\epsilon_{\alpha \beta} \partial_{\alpha} A_{\beta}^{(\msf{orb})} = B_z$.

Non-magnetic 
adatoms or charge traps are encoded in the 
scalar potential $V(\vex{r})$. In-plane (out-of-plane) polarized magnetic 
impurities additionally induce point exchange coupling to the vector $\vex{A}(\vex{r})$ 
[mass $M(\vex{r})$] fields.\cite{SingImpRef} 
The different types of disorder leading to $V$, $\vex{A}$, and $M$ are sketched
in Fig.~\ref{DirtVar}.
Assuming a random surface distribution of impurities
and spatial rotational invariance on average,
the disorder potentials can be taken as Gaussian white noise distributed
variables,
\begin{align}\label{WhiteNoise}
\begin{aligned}
	\overline{V(\vex{r}) V(\vex{r'})} &= \Delta_{V} \, \vf^2 \, \delta(\vex{r} - \vex{r'}), \\
	\overline{A_\alpha(\vex{r}) A_\beta(\vex{r'})} &= \Delta_A \, \vf^2 \, \delta_{\alpha \beta} \, \delta(\vex{r} - \vex{r'}), \\
	\overline{M(\vex{r}) M(\vex{r'})} &= \Delta_M \, \vf^2 \, \delta(\vex{r} - \vex{r'}).
\end{aligned}
\end{align}
The dimensionless variances $\Delta_{V,A,M}$ quantify the disorder strength.
In the first Born approximation, these are of the form
\begin{align}\label{DeltaMicro}
	\Delta \, \vf^2 = \nimp |\tilde{u}(0)|^2,
\end{align}
where $\nimp$ is the impurity density, and $\tilde{u}(\vex{q})$ denotes the Fourier transform 
of the single impurity potential. 
We note that a net in-plane magnetization of the surface impurities 
$\overline{A^\mu} \neq 0$ can be removed by a gauge transformation, while the average scalar 
potential $\overline{V}$ is absorbed into the chemical potential. We will assume that there is no net
magnetization perpendicular to the surface, $\overline{M} = 0$, or that we only
probe LDOS fluctuations on energy scales much larger than the induced gap $2 \vf \overline{M}$.

In 2D, the single impurity potential $u(\vex{r})$ [Eq.~(\ref{DeltaMicro})] must 
decay faster than $1/r^2$ (or oscillate 
rapidly enough)
so that the limit $\tilde{u}(\vex{q}\rightarrow 0)$ exists; 
otherwise, the white noise assumption in Eq.~(\ref{WhiteNoise}) is invalidated by long range impurity 
potential correlations.\cite{LRCorrDirt} This causes a problem for charged impurities, which can 
become poorly screened for a small surface doping relative to the Dirac point. In graphene, the 
long-range correlated potential undulations induced by 
poorly-screened 
substrate impurities leads to 
a smearing of the Dirac point over an energy scale 
$k_B T_{\msf{rms}} \propto \vf \sqrt{\nimp}$, 
and to the breakup of the sample
into electron and hole puddles.\cite{Puddles,PuddlesReview} The advent of electron-hole
puddles has until recently prevented the observation of various ``intrinsic'' phenomena associated
to the Dirac carriers in graphene experiments such as velocity renormalization\cite{PuddlesNoMore}
and hydrodynamic transport near the Dirac point. In this respect, a \emph{large} surface or 
bulk doping actually improves the situation for STM measurement of disorder-induced quantum 
interference, since these carriers screen the potential of surface charges. The disorder potential
can be considered short-range correlated for scales larger than the screening length. 

If we consider only surface doping, with an insulating bulk, then the Thomas-Fermi
wavelength due to a finite surface carrier density $n$ is given by
\begin{align}
	\lambda_{\msf{TF}}
	=
	\frac{1}{\alpha}
	\sqrt{\frac{\pi}{n}},
\end{align}	
where $\alpha \equiv e^2 / \epsilon \vf$ is the effective ``fine structure constant.''
The permittivity $\epsilon = (1 + \epsilon_{\text{TI}})/2$, the average of the
bulk TI below and vacuum above the surface.
For $\text{Bi}_2 \text{Se}_3$ with a surface density of 
$n = 7 \times 10^{12}$ $\text{cm}^{-2}$ (corresponding to 
a doping level of $0.3$ eV relative to the Dirac point),\cite{Bi2X3 BS} 
$\vf = 5 \times 10^5$ m/s (Ref.~\onlinecite{Bi2X3 BS}), 
and permittivity\cite{Bi2Se3Diel} $\epsilon_{\text{TI}} = 113$,
one obtains $\lambda_{\msf{TF}} \sim 90$ nm.
This is very large, and indicates that the surface state carrier
density is inadequate to screen charged impurities.
A smaller screening length is possible for bulk doping,\cite{Yazdani11}
or by performing experiments on thin film samples exfoliated over 
a metallic gate. Alternatively, one can restrict the 
deposition of surface impurities to non-doping adatoms,
e.g.\ iron in $\text{Bi}_2 \text{Se}_3$.\cite{Shen10}  
The disorder variance associated to Thomas-Fermi screened charged impurities is 
\begin{align}\label{Delta_v Screened}
	\Delta_{V} = \pi \frac{\nimp}{n}. 
\end{align}	

Finally, we note that the appearance in isolation of any of
the three disorder potentials in Eq.~(\ref{H}) realizes
three different symmetry classes of Anderson (de)localization,\cite{Classes,BernardLeClair,SRAL08}
see Appendix~\ref{App--Sym} for a review.
The $\mathcal{T}$-invariant case with $\Delta_{A,M} = 0$ belongs
to the spin-orbit class AII, which is also the class of the
$\mathbb{Z}_2$ topological bulk 
[Fig.~\ref{DirtVar}{\bf (a)}].
In the case of broken $\mathcal{T}$,
$\Delta_{V,M} = 0$ realizes 
the random vector potential model in 
class AIII
[Fig.~\ref{DirtVar}{\bf (b)}],
while $\Delta_{V,A} = 0$ gives 
the random mass model in
class D
[Fig.~\ref{DirtVar}{\bf (c)}].
All three classes exhibit delocalized states in 2D,
although this occurs only at the Dirac point for class AIII.\cite{Ludwig94}
In the $\mathcal{T}$-invariant symplectic case, the unpaired single Dirac
flavor avoids the usual spin-orbit metal-insulator transition,\cite{Z2Numerics}
remaining delocalized even for strong disorder due to a topological term.\cite{Z2Shinsei}
The generic case of broken-$\mathcal{T}$ with all three disorder
potentials non-zero realizes the unitary class A, and is believed
to flow under renormalization to the plateau transition in the 
integer quantum Hall effect.\cite{Ludwig94,NomuraRyu08} 
(See Sec.~\ref{IQHP} for a review).

Because in-plane (out-of-plane) Zeeman coupling appears in the
vector (mass) potential [Eq.~(\ref{VectorMass})], one is tempted
to identify class AIII (class D) with the limit of an otherwise
clean surface, dusted with charge neutral magnetic impurities 
randomly polarized in-plane (perpendicular to the TI surface). 
However, a magnetic adatom is expected to also induce
a local scalar potential deformation $V(\vex{r})$.
For example, 
it can dope the surface or bulk,
as occurs for a manganese impurity in $\text{Bi}_2 \text{Se}_3$
(Ref.~\onlinecite{Shen10})].
As discussed
in Appendix~\ref{App--Sym}, the advent of any two flavors of disorder
destroys the additional discrete symmetries enjoyed by
the special class D and AIII Hamiltonians. The asymptotic long-distance
LDOS scaling is then governed by the unitary class A, discussed above.
Nevertheless, depending upon the relative microscopic strength of the magnetic versus 
potential perturbations induced by polarized magnetic impurities, the class AIII or D
model may provide an adequate approximation for 
broken-$\mathcal{T}$ 
LDOS fluctuations on intermediate scales.

\subsection{Results\label{Results}}

To compute the scaling of LDOS moments in a quantum theory with
quenched disorder, one employs a path integral $Z$ to express products of 
fermion Green's functions as functionals of the disorder configuration. Using a
trick (replicas,\cite{Wegner,LeeRamakrishnan,Boris} supersymmetry,\cite{SUSY} 
or  Keldysh\cite{Keldysh}) to normalize $Z = 1$, the Green's functions are
formally averaged over disorder configurations (typically with a Gaussian weight). 
The result is a translationally-invariant, but ``interacting'' field theory, where 
the disorder strength $\Delta$ appears as a coupling constant.\cite{LeeRamakrishnan,SUSY} 
Perturbative calculations are controlled via loop expansion for small $\Delta$. 

To determine the scaling, one decomposes the $q^{\text{th}}$ LDOS moment into projections
upon the renormalization group (RG) eigenoperators of the disorder-averaged theory.\cite{Wegner,Pruisken,Boris} 
The multifractal spectrum $\tau(q)$ is determined by the most relevant (negative)\cite{Duplantier} 
scaling dimension $x_q$ exhibited by an eigenoperator in this decomposition,
and is given by\cite{Mirlin,Foster09} 
\begin{align}\label{tauScalingDim}
	\tau(q) = 2(q - 1) + x_q - q x_1.
\end{align}

\subsubsection{Broken $\mathcal{T}$: random vector potential disorder (Class AIII)}

The properties of the model in Eq.~(\ref{H}) with short-range correlated
disorder [Eq.~(\ref{WhiteNoise})] were originally studied in Ref.~\onlinecite{Ludwig94}.
In this work, the exact multifractal spectrum $\tau(q)$ was calculated for the broken-$\mathcal{T}$,
random vector potential ($\sim$ in-plane polarized magnetic 
impurity)\cite{Footnote--IsolatedDisorder} 
class AIII model, 
to all orders in $\Delta_A$. Technically, this result obtains because the disorder-averaged
AIII model is conformally invariant at the Dirac point, and the exact LDOS moment spectra can be extracted
using an Abelian bosonization treatment. The exact spectrum\cite{Ludwig94} is quadratic in $q$, and takes the 
form of Eq.~(\ref{tauQuad}), with 
\begin{align}\label{tauAIII}
	\theta_A = \frac{\Delta_A}{\pi}.
\end{align} 
Subsequent work\cite{Chamon96,Castillo97} on the random vector potential model elucidated 
the mechanisms of 
termination and freezing, transitions that occur in the 
spectral statistics for large moments $q > q_c(\Delta_A)$ or strong disorder 
$\Delta_A \geq 2 \pi$. 

For this broken-$\mathcal{T}$ class, we can also examine the spin LDOS
fluctuations, utilizing the same nonperturbative bosonization treatment employed in 
Ref.~\onlinecite{Ludwig94}. The spin LDOS $\nu^{\hat{\iota}}(\varepsilon,\vex{r})$ taken along
an axis $\hat{\iota}$ in spin space was defined by Eq.~(\ref{SpinLDOS}).
Moment fluctuations are characterized by the inverse spin participation ratio 
(ISPR) in Eq.~(\ref{ISPR}), the scaling limit of which is controlled by the
dimension $x_{q}^{\hat{\iota}}$ that appears in Eq.~(\ref{ISPR--Scaling}). 
The out-of-plane ISPR $\mathcal{P}^{\hat{3}}_{q}(\varepsilon)$ is associated
to the ``mass'' fermion bilinear $\nu^{\hat{3}} = \psi^\dagger \sigh_3 \psi$.
For the random vector potential model, the most relevant contribution to
$\mathcal{P}^{\hat{3}}_{q}(\varepsilon)$ carries the same scaling dimension that 
gives the composite LDOS scaling in Eqs.~(\ref{tauQuad}) and (\ref{tauAIII}),
\begin{align}\label{x_q^3 AIII}
	x_{q}^{\hat{3}} = q - \frac{\Delta_A}{\pi} q^2.
\end{align}
The chiral components of the in-plane spin LDOS are the energy-resolved $U(1)$ Dirac current
operators
\begin{align}\label{spinLDOSChipmDef}
	\nu^{\pm} \equiv \nu^{\hat{1}} \pm i \nu^{\hat{2}} = \psi^\dagger \sigh_{\pm} \psi.
\end{align}
Moments 
of these are RG eigenoperators that receive no corrections.
The scaling of the associated ISPR is governed by the disorder-independent
(tree level) exponent
\begin{align}\label{x_q^pm AIII}
	x_{q}^{\pm} = q.
\end{align}
Eqs.~(\ref{tauAIII}), (\ref{x_q^3 AIII}), and (\ref{x_q^pm AIII}) are exact results that hold to all
orders in $\Delta_A$.

\subsubsection{Broken $\mathcal{T}$: random mass disorder (Class D)}

In the rest of this section, we provide new results for the broken-$\mathcal{T}$, random mass 
($\sim$ out-of-plane polarized magnetic 
impurity)\cite{Footnote--IsolatedDisorder} 
class D model,
the $\mathcal{T}$-invariant class AII model,
and the generic broken-$\mathcal{T}$ unitary class A model.
For weak disorder, 
none of these are
conformally invariant, and we resort to perturbation 
theory. In this section we summarize results; some technical aspects
are sketched in Appendix~\ref{App--PT}. The results obtained below
hold only for small $\Delta_{V,M} \ll 1$, wherein the disorder 
appears as a weak marginal perturbation (at tree level) to the 
clean Dirac surface band structure.

For the broken-$\mathcal{T}$ case of random mass disorder 
(with $\Delta_{V} = \Delta_A = 0$),
one obtains quadratic multifractality at one loop,
again governed by Eq.~(\ref{tauQuad}), with
\begin{align}\label{tauD}
	\theta_M = \frac{\Delta_M}{2 \pi} + \ord{\Delta_M^2}.
\end{align}
Moments of the out-of-plane spin LDOS operator $\nu^{\hat{3}} = \psi^\dagger \sigh_3 \psi$,
as well as of the chiral in-plane [$U(1)$ current] operators $\nu^{\pm} = \psi^\dagger \sigh_{\pm} \psi$
constitute RG eigenoperators at one loop, with scaling dimensions
\begin{align}\label{x_q^3 D}
	x_{q}^{\hat{3}} =& q + \frac{\Delta_M}{2 \pi} q + \ord{\Delta_M^2},
	\\
	\label{x_q^pm D}
	x_{q}^{\pm} =& q + \ord{\Delta_M^2}.
\end{align}
Note that the first correction in Eq.~(\ref{x_q^3 D}) is \emph{positive}
(and linear in $q$);
this should be contrasted with the AIII case, Eq.~(\ref{x_q^3 AIII}) above. 
On general grounds, the anomalous scaling dimension associated
to the $q^{\text{th}} \geq 1$ moment of the composite LDOS, or any projected component 
thereof, must appear with a negative sign. The reason is that this quantity 
is associated to a moment of a normalized probability distribution\cite{MFCReview,Duplantier} 
through Eqs.~(\ref{Box Prob}) and (\ref{IPR}).
For a quadratic $\tau(q)$ spectrum, this leads in particular to $\theta \geq 0$ 
in Eq.~(\ref{tauQuad}) [consistent with a positive, real disorder variance---c.f.\
Eqs.~(\ref{tauAIII}), (\ref{tauD}), and (\ref{tauAII})]. By contrast, the spin LDOS is defined 
as the \emph{difference} between two orthogonal projections [Eq.~(\ref{SpinLDOS})]; 
for this reason, the first disorder correction to the scaling dimension 
in Eq.~(\ref{x_q^3 D}) is not required to appear with a particular sign.

\subsubsection{Non-magnetic disorder (Class AII)}

In the $\mathcal{T}$-invariant case of scalar potential disorder, 
it turns out that no local operator (without derivatives) exhibits multifractal scaling to
first order in $\Delta_{V}$. For Dirac fermions, this applies
to both LDOS and energy-resolved current moments.
Physically, the weak influence of non-magnetic disorder is due to 
interference mediated by the Dirac pseudospin (equivalent to physical spin $1/2$ on the TI surface).
The Dirac pseudospin is also responsible for the suppression of backscattering 
from a single non-magnetic impurity.\cite{SuppBackScat}
Technically, this result is derived by mapping the one-loop renormalization 
process of local operators to the action of a certain spin-$1/2$ 
Hamiltonian $H^{(\msf{eff})}_V$, and identifying renormalization 
group eigenoperators with states that diagonalize $H^{(\msf{eff})}_V$ 
(see Appendix~\ref{App--PT}). As a result, to lowest order one observes 
plane wave scaling in the LDOS IPR [Eq.~(\ref{IPR})]. The spin LDOS
vanishes exactly, due to $\mathcal{T}$. 

The first non-trivial correction to the LDOS $\tau(q)$ appears at two loops. 
To this order, the spectrum is again quadratic as in Eq.~(\ref{tauQuad}).
A straight-forward but laborious calculation gives the coefficient in this equation,
\begin{align}\label{tauAII}
	\theta_V
	= 
	\frac{3 \Delta_{V}^2}{8 \pi^2}
	+ \ord{\Delta_{V}^3}.
\end{align}
Since $\Delta_{V} \propto \nimp$ [Eqs.~(\ref{DeltaMicro}) or (\ref{Delta_v Screened})], 
we find that the non-trivial multifractal scaling begins at second order in the impurity density. 
This is qualitatively weaker than \emph{any} of the 
broken-$\mathcal{T}$ 
regimes, where
the quadratic multifractality appears already at first order, Eqs.~(\ref{tauAIII}) and (\ref{tauD}).
This distinction between $\mathcal{T}$-invariant and $\mathcal{T}$-broken surfaces
is our primary result, and can be tested directly in STM experiments by varying the concentration
of deposited surface disorder.  
Although the $\mathcal{T}$-invariant case is not conformally invariant
(for a discussion of renormalization effects, see Sec.~\ref{Renorm}, below),
the multifractal $\tau(q)$ and $f(\alpha)$ spectra depend only upon a single
parameter, the variance $\Delta_{V}$. Eq.~(\ref{tauAII}) can be extended
to higher loops, allowing ever more precise tests against numerics or 
experimental data within the perturbatively accessible regime.
The multifractal spectrum therefore provides a unique fingerprint for the 
time-reversal invariant Dirac surface state of the $\mathbb{Z}_2$ topological
insulator, in the presence of 
weak but otherwise
generic non-magnetic disorder.
The opposite limit of strong disorder for the $\mathcal{T}$-invariant
case is discussed below in Sec.~\ref{SympMetal}.

\begin{figure}
   \includegraphics[width=0.45\textwidth]{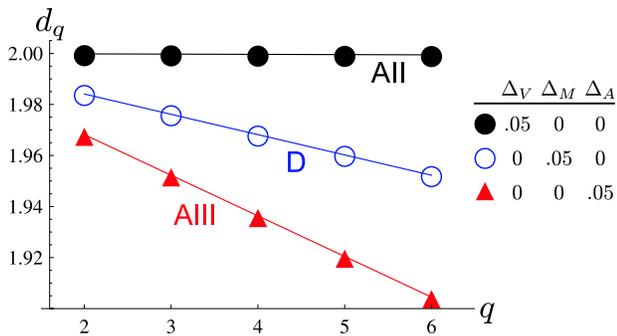}
   \caption{
	Quadratic multifractality for isolated disorder flavors.
	The Renyi dimension $d_q = 2 - \theta \, q$ 
	[Eqs.~(\ref{tauQuad}) and (\ref{Renyi})]
	is plotted for the exact vector potential (AIII),
	one-loop mass (D),
	and two-loop scalar potential (symplectic AII)
	results [Eqs.~(\ref{tauAIII}), (\ref{tauD}), and (\ref{tauAII})].
	The disorder strength is $\Delta = 0.05$ for each case. 
	The broken time-reversal class D and AIII corrections
	appear at order $\Delta_{M,A}$, while the 
	(much weaker) time-reversal invariant class AII correction
	begins at order $\Delta_V^2$.
	}
   \label{FigRenyi1}
\end{figure}

\subsubsection{Broken $\mathcal{T}$: generic disorder (Class A)\label{ClassA}}

\begin{figure}[b]
   \includegraphics[width=0.45\textwidth]{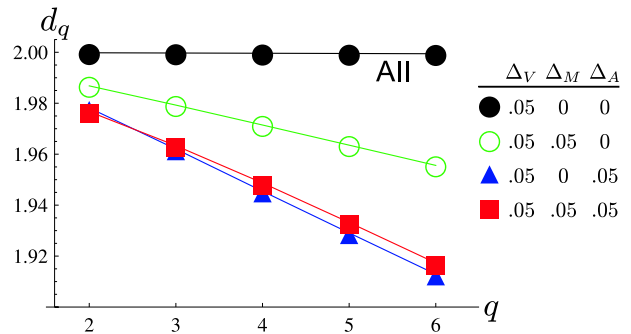}
   \caption{
	One-loop Renyi dimensions [Eq.~(\ref{Renyi})] 
	in the broken $\mathcal{T}$, multidisorder
	unitary class A case, for various disorder strength
	combinations. These results were obtained by numerically 
	extracting the largest positive eigenvalue from the effective 
	spin Hamiltonian $H^{\msf{eff}}$ in Eq.~(\ref{HeffSummed})
	(restricting the search to operators invariant under
	spatial rotations and reflections);
	see Appendix~\ref{App--PT} for details.
	The two-loop result for the $\mathcal{T}$-invariant case
	is shown for reference.
	}
   \label{FigRenyi2}
\end{figure}

When $\mathcal{T}$ is broken and any two disorder flavors appear,
the system resides in the unitary class A. The third disorder
flavor is always generated under renormalization---see Sec.~\ref{Renorm},
below.  The results of Eqs.~(\ref{tauAIII}) and (\ref{tauD}) for the 
LDOS $\tau(q)$ spectrum in the random vector and 
mass potential models suggest that the 
unitary 
case 
also exhibits
multifractality to first order in the impurity density $\nimp$, since $\Delta_{A,M,V} \propto \nimp$. 

With multiple flavors of the disorder,
solving the operator mixing problem for the $q^{\text{th}}$ 
LDOS moment requires the diagonalization of an effective spin Hamiltonian
$H^{(\msf{eff})}$, transcribed in Eq.~(\ref{HeffSummed}) of
Appendix~\ref{App--PT}. 
In Figs.~\ref{FigRenyi2} and \ref{FigRenyi3},
we present the results obtained by numerically diagonalizing this 
matrix for various combinations of $\{\Delta_V,\Delta_M,\Delta_A\}$.
In these figures we plot the  
\emph{Renyi dimension}\cite{PaladinVulpiani} 
$d_q$,
defined for $q \neq 1$ via
\begin{align}\label{Renyi}
	d_q 
	\equiv 
	\frac{\tau(q)}{q - 1}.
\end{align}
Figs.~\ref{FigRenyi2} and \ref{FigRenyi3} show that the generic
broken-$\mathcal{T}$
case is multifractal at one loop, and easily distinguished from
the two-loop $\mathcal{T}$-invariant result, in the limit of weak disorder. 
[Note that Fig.~\ref{FigRenyi3} indicates that the $\tau(q)$ spectrum is not purely
quadratic in this general case.]
It should therefore be 
possible to precisely distinguish the broken-$\mathcal{T}$ and 
$\mathcal{T}$-invariant spectra experimentally, by observing the dependence of 
the deviation $2 - d_q$ on $\nimp$. 
The single-disorder flavor results for comparable strengths are plotted
in Fig.~\ref{FigRenyi1} for reference.

For the multidisorder unitary model,
the same RG eigenoperators dominate the scaling of composite $\nu$ and 
out-of-plane spin $\nu^{\hat{3}}$ LDOS moments. The dimension 
$x_q^{\hat{3}}$ that determines the spin LDOS scaling via Eq.~(\ref{ISPR--Scaling}) 
also
enters into the LDOS spectrum in Eq.~(\ref{tauScalingDim}), leading to Figs.~\ref{FigRenyi2} 
and \ref{FigRenyi3}.
By contrast, moments of the chiral spin LDOS 
components $\nu^{\pm}$ [Eq.~(\ref{spinLDOSChipmDef})] 
remain eigenoperators that acquire no corrections at one loop,
\begin{align}\label{x_q^pm A}
	x_{q}^{\pm} = q
	+
	\ord{\Delta_{\alpha} \Delta_{\beta}},
\end{align}
$\alpha,\beta \in \{A,M,V\}$.

As reviewed in 
the subsequent
Sec.~\ref{IQHP}, 
for $\overline{M} = 0$, the
generic 
broken-$\mathcal{T}$ 
model is believed to flow to the critical
state at the integer quantum Hall plateau transition.\cite{Ludwig94,NomuraRyu08}
This state exhibits strong multifractality that has been extensively
studied in 
numerics.\cite{MFCReview,Mirlin,PookJanssen91,KlesseEvers95,ObuseEvers08}

\begin{figure}
   \includegraphics[width=0.45\textwidth]{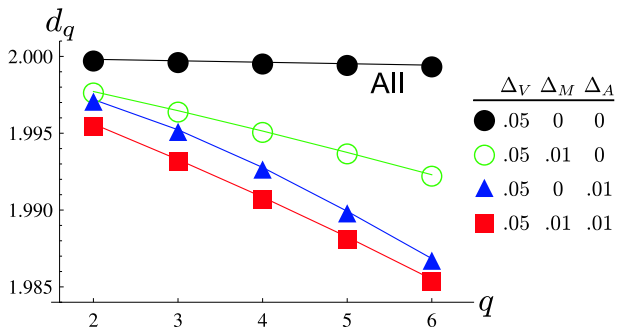}
   \caption{
	The same as Fig.~\ref{FigRenyi2}, but for different unitary class
	disorder strength combinations.
	In
	the data presented here, $\Delta_V > \Delta_{M,A}$.
	Regardless, the one-loop spectrum obtained for 
	either $\Delta_{M,A} > 0$ is stronger than the two loop 
	$\mathcal{T}$-invariant case with $\Delta_M = \Delta_A = 0$. 
	The latter is also shown for reference.
	}
   \label{FigRenyi3}
\end{figure}


\subsection{Renormalization effects \label{Renorm}}

As discussed at the beginning of the previous section,
the disorder-averaged Dirac surface state theory 
used to compute LDOS multifractal spectra is
an ``interacting'' field theory, wherein the disorder
strengths $\Delta_{V,A,M}$ appear as coupling constants
(c.f.\ Appendix \ref{App--PT}). 
Because
these parameters
are dimensionless, at weak coupling the disorder constitutes
a marginal perturbation of the clean Dirac band structure. 
The one-loop RG equations for these parameters are given 
by\cite{Ludwig94,Mirlin1loopRG} 
\bsub\label{1loopRG}
\begin{align}
	\frac{d \Delta_A}{d l} &= \frac{1}{\pi} \Delta_M \Delta_{V},
	\\
	\frac{d \Delta_M}{d l} &= \frac{1}{\pi} \left(2 \Delta_A - \Delta_M \right)\left(\Delta_M + \Delta_{V}\right),
	\\
	\frac{d \Delta_{V}}{d l} &= \frac{1}{\pi} \left(2 \Delta_A + \Delta_{V} \right)\left(\Delta_M + \Delta_{V}\right),
\end{align}
\esub
where $l = \log L$ denotes the log of the RG length scale (e.g., the system size).
Energy $\varepsilon$ scales as 
\begin{align}\label{DynScaling}
	\frac{d \ln \varepsilon}{d l} = z(l),
\end{align}
where the (scale-dependent) dynamic critical exponent is 
\begin{align}\label{DynCritExp}
	z = 1 + \frac{1}{2 \pi}\left(2 \Delta_A + \Delta_M + \Delta_{V}\right)
	+
	\ord{\Delta_{\alpha} \Delta_{\beta}},
\end{align}
$\alpha,\beta \in \{A,M,V\}$.

In this section, we use Eqs.~(\ref{1loopRG})--(\ref{DynCritExp}) to derive 
the dynamical scaling of the disorder parameters $\Delta_{V,A,M}(\varepsilon)$;
here energy $\varepsilon$ is measured relative to the Dirac point, \emph{not} the Fermi energy. 
(From the point-of-view of the disordered Dirac theory, a finite energy above the Dirac
point constitutes a relevant perturbation.\cite{Ludwig94})
Using the results obtained in the previous section, we thereby determine the 
enhancement or suppression of LDOS multifractality approaching the Dirac point, 
due to renormalization.

\subsubsection{Broken $\mathcal{T}$: random vector potential disorder (Class AIII)}

For the random vector potential model with $\Delta_{V} = \Delta_M = 0$,
Eq.~(\ref{1loopRG}) implies
\[
	\frac{d \Delta_A}{d l} = 0,
\]
so that $\Delta_A = \dAo$ (constant), where $\dAo$ is the ``microscopic'' value derived
from the randomly polarized in-plane magnetic impurity distribution.\cite{Footnote--IsolatedDisorder}  
This result in fact holds to all orders in $\Delta_A$;\cite{Ludwig94} in this case, the theory describing LDOS
fluctuations at the Dirac point is conformally invariant.
Multifractality is neither enhanced nor suppressed as one moves away
from the Dirac point, defined as $\varepsilon = 0$. However, for non-zero 
energies $\varepsilon \neq 0$, in an infinite size sample all states are in fact localized.\cite{Ludwig94} 
The localization length diverges upon approaching the band center as $\loc(\varepsilon) \sim \varepsilon^{-1/z}$, with 
$z = 1 + \Delta_A/\pi$ [Eq.~(\ref{DynCritExp})]. Eqs.~(\ref{tauQuad}) and (\ref{tauAIII}) for 
$\tau(q)$ hold on scales smaller than $\loc(\varepsilon)$.

\subsubsection{Broken $\mathcal{T}$: random mass disorder (Class D)}

For the random mass model with $\Delta_{V} = \Delta_A = 0$,
\[
	\frac{d \Delta_M}{d l} = - \frac{\Delta_M^2}{\pi} + \ord{\Delta_M^3},
\]
so that the disorder is marginally irrelevant at weak coupling.\cite{Ludwig94}
Integrating this equation and using Eqs.~(\ref{DynScaling}) and (\ref{DynCritExp}),
we can compute the scaling of $\Delta_M$ with energy. 
At energy scale $\Upsilon$, we define $\dmo \equiv \Delta_M(\Upsilon)$; 
then for the smaller energy scale $\varepsilon$
(relative to the Dirac point),
we obtain the logarithmic suppression
\begin{align}\label{D: DynScaling pert}
	\Delta_M(\varepsilon \lesssim \Upsilon)
	\sim&
	\,
	\dmo
	-
	\frac{\left(\dmo\right)^2}{\pi} 
	\log\left(\frac{\Upsilon}{\varepsilon}\right)
	\nonumber\\
	&+
	{\bm{\mathit{O}}}\left\{
	\left[\dmo\left(1 - \frac{\varepsilon}{\Upsilon}\right)\right]^2\!,
	\left(\dmo\right)^3
	\right\}\!.
\end{align}
This equation holds for $| 1 - \varepsilon/\Upsilon | \ll 1$.
In the limit as $\varepsilon \rightarrow 0$, the disorder
strength vanishes as
\begin{align}\label{D: DynScaling asym}
	\Delta_M(\varepsilon \rightarrow 0)
	\sim&
	\pi 
	\left[
	\ln
	\left(
	\sqrt{\frac{\pi}{\dmo}}
	\frac{\Upsilon}{\varepsilon}
	\right)
	\right]^{-1}
	\nonumber\\
	&+
	\sqord{
	\frac{1}{\dmo} 
	\ln^{-2}
	\left(
	\sqrt{\frac{\pi}{\dmo}}
	\frac{\Upsilon}{\varepsilon}
	\right)
	}.
\end{align}
For small $\dmo$, Eq.~(\ref{D: DynScaling asym}) applies
only at very small energies $\varepsilon \lesssim \Upsilon \exp(-1/\dmo)$.

\subsubsection{Non-magnetic disorder (Class AII)}

Now we consider the $\mathcal{T}$-invariant model.
The flow equation for $\Delta_{V}$ is 
\begin{align}\label{AIIFlow}
	\frac{d \Delta_{V}}{d l} = \frac{\Delta_{V}^2}{\pi} + \ord{\Delta_{V}^3}.
\end{align}
In contrast to the random mass, the random scalar potential
is a marginally relevant perturbation to the clean
band structure.\cite{Ludwig94} Examining lower and 
lower energy scales approaching the Dirac point, one
observes stronger effects of the disorder. In the asymptotic
scaling limit wherein the 
impurity potential
strength becomes ``large''
($\Delta_{V} \gtrsim 1$), numerical\cite{Z2Numerics} results and analytical\cite{Z2Shinsei}
arguments imply that the disordered $\mathcal{T}$-invariant Dirac theory
renormalizes into the ``conventional'' symplectic metal.
The metal is distinguished from the Dirac theory by its non-zero 
(and non-critical) density of states at zero energy,\cite{LeeRamakrishnan}
and by its $\tau(q)$ spectrum.\cite{Wegner,FalkoEfetov95}  
We discuss the strong coupling LDOS multifractality 
below
in Sec.~\ref{SympMetal}. 
If at energy $\Upsilon$, $\Delta_{V}(\Upsilon) \equiv \dvo \ll 1$,
then for a somewhat smaller energy $\varepsilon$ we obtain the logarithmic
enhancement
\begin{align}\label{AII: DynScaling pert}
	\Delta_{V}(\varepsilon \lesssim \Upsilon)
	\sim&
	\dvo 
	+
	\frac{\left(\dvo \right)^2}{\pi}
	\log\left(\frac{\Upsilon}{\varepsilon}\right)
	\nonumber\\
	&+
	{\bm{\mathit{O}}}\left\{
	\left[\dvo\left(1 - \frac{\varepsilon}{\Upsilon}\right)\right]^2\!,
	\left(\dvo\right)^3
	\right\}\!.
\end{align}

Eq.~(\ref{AII: DynScaling pert}) implies that renormalization strengthens multifractality
approaching the Dirac point $\varepsilon = 0$, for the $\mathcal{T}$-invariant case.  
We emphasize that this has \emph{nothing} to do with weak (anti-)localization.
The latter occurs in the diffusive metallic regime with $k_{F} l_{\msf{mfp}} \gg 1$, 
where $l_{\msf{mfp}}$ denotes the elastic mean free path. 
The diffusive regime obtains at \emph{strong}
coupling\cite{Z2Numerics} 
near the Dirac point 
$k_F \rightarrow 0$ [Sec.~\ref{SympMetal}, below]. 
The impurity strength renormalization in Eq.~(\ref{AII: DynScaling pert}) is a 
quantum effect deriving from the clean band structure, in the ``near-ballistic'' regime.\cite{Schuessler09}

\subsubsection{Broken $\mathcal{T}$: generic disorder (Class A)\label{ClassA Ren}}

In the generic case of broken $\mathcal{T}$, with multiple disorder
coupling strengths non-zero, the system flows 
toward
strong coupling
$\Delta_{V,M,A} \rightarrow \infty$. As a result, multifractality
is enhanced approaching the Dirac point.
The RG flow ultimately terminates at a strong coupling critical point, or
in the Anderson insulator, discussed in the next section.

\section{Strong coupling regimes \label{Sec: Strong}}

In this section, we review prior results on strong coupling
regimes relevant to the disordered Dirac $\mathbb{Z}_2$
topological insulator surface states, LDOS fluctuations and
associated multifractal spectra. These are not new, but
provide complimentary information to 
the new
results derived in the previous section.  

In both generic cases of $\mathcal{T}$-invariant, and
$\mathcal{T}$-breaking impurities, the disordered
Dirac description used in Sec.~\ref{Sec: Weak}
fails on the largest length and lowest energy scales
(approaching charge neutrality). 
For a sufficiently dilute concentration impurities, 
the results obtained in the previous section
characterize the start of the scaling regime,
over energy and length scales such that 
the disorder strengths remain weak $\Delta_{V,A,M}(L,\varepsilon) \ll 1$.
When these parameters become order one 
(due to renormalization down to lower energies and longer lengths), 
the system crosses
over to one of the strong coupling regimes discussed below.

\subsection{Delocalized states at strong disorder}

\subsubsection{$\mathcal{T}$-invariant case: diffusive metal via strong disorder \label{SympMetal}}

In a random scalar potential field, the Dirac point 
vacillates in energy with spatial location; as a result, the density of states
near charge neutrality is enhanced by the disorder. Due to the suppression of
pure backscattering for Dirac 
fermions,\cite{SuppBackScat} 
the state density enhancement more than compensates for the increased scattering 
introduced by the additional impurities. As a result, scalar potential disorder actually 
\emph{increases} the (zero temperature, Landauer) conductance at charge 
neutrality beyond the clean ballistic result, $e^2/\pi h$
(Refs.~\onlinecite{Z2Numerics,Schuessler09}).
As 
in Sec.~\ref{Sec: Weak},
here we assume short-range correlated disorder, due either
to charge neutral impurities or efficient screening by bulk and/or surface carriers.
We do not discuss the puddle regime\cite{Puddles,PuddlesReview} in the present
paper.

The effective disorder strength $\Delta_V$ is enhanced by renormalization,
as indicated by the runaway flow implied by Eq.~(\ref{AIIFlow}). 
The concomitant density of states and conductance growth
suggests that the disordered Dirac theory ultimately crosses over
to the ordinary diffusive symplectic metal, a result born out 
by numerics.\cite{Z2Numerics} In the absence of time-reversal 
symmetry breaking, Anderson localization is prohibited on the surface 
of a topological insulator.\cite{SRAL08,Z2Shinsei}

The symplectic metal possesses a finite (non-critical) average density of states
at charge neutrality, and a distinct $\tau(q)$ spectrum. For a large
effective diffusion constant $D$ (induced for a Dirac fermion subject to sufficiently 
\emph{strong} 
disorder,\cite{Z2Numerics}
or for a weakly disordered system examined on 
large length scales), the lowest order result for the multifractal spectrum appears 
in Eqs.~(\ref{tauQuad}) and Eq.~(\ref{tauMetal}), above. In the latter equation, 
$\beta = 4$ for the symplectic class.\cite{FalkoEfetov95,RandomMatrix} 

For the $\mathcal{T}$-invariant case, the strongest multifractality is
expected at intermediate coupling. Weak disorder $\Delta_V \ll 1$ induces 
weak multifractality in the Dirac language [$\tau(q)$ in Eqs.~(\ref{tauQuad}) and (\ref{tauAII})], 
while strong disorder ultimately pushes the system into the symplectic metal, where
a large diffusion constant $D$ suppresses the first correction in Eq.~(\ref{tauMetal}).

\subsubsection{Broken $\mathcal{T}$: IQHP transition\label{IQHP}}

For generic $\mathcal{T}$-breaking disorder, i.e.\ all three 
$\Delta_{V,M,A}$ non-zero, the disordered Dirac theory is 
also
unstable under renormalization. 
When the average mass is zero 
$\overline{M} = 0$ (see below), the flow in Eq.~(\ref{1loopRG}) is 
believed to terminate at the critical point of the integer quantum Hall plateau transition.\cite{Ludwig94,NomuraRyu08} 
This is the delocalized state separating adjacent Hall plateaux; it exhibits \emph{strong} multifractality 
that has been extensively studied in numerics.\cite{MFCReview,PookJanssen91,KlesseEvers95,ObuseEvers08}
The spectrum is believed to be universal,\cite{MFCReview} and is approximately\cite{ObuseEvers08} 
parabolic as in Eqs.~(\ref{tauQuad}) and (\ref{fPara}), with $\theta \sim 0.26$
(Refs.~\onlinecite{KlesseEvers95,ObuseEvers08}).

\subsection{Anderson insulator \label{Sec: Anderson Ins}}

At zero chemical potential relative to the Dirac point, an average
out-of-plane spin magnetization at the surface of a 
$\mathbb{Z}_2$
TI corresponds to the presence of a non-zero Dirac mass $M$ for the
surface carriers. This insulating state resides in a quantum Hall 
plateau [with $\sigma_{xy} = \sgn(M) \, e^2/2 h$].\cite{Haldane88,Ludwig94,KaneHasanREVIEW,ZhangREVIEW}
In the presence of surface disorder, the plateau state will
assume the character of a localized Anderson insulator. 
In this section we review LDOS fluctuations in the Anderson insulator. 
The discussion is relevant not only to the magnetized surface of
a 3D TI, but also to 
localized states populating the bulk gap of a 
disordered TI.
Proposals
exploiting localization to realize 
so-called ``topological Anderson insulators'' 
by adding impurities to clean 
hosts 
include 
those in 
Ref.~\onlinecite{TopAndIns}.

To understand local density of states fluctuations in an 
Anderson insulator, it is useful to first 
study
a toy
problem.
Consider a tight-binding model on a $d-$dimensional
lattice, subject to nearest-neighbor hopping $t$, and 
a random on-site potential $V_i$, distributed uniformly
over the region $-W/2 \leq V_i \leq W/2$. We assume the
absence of spatial correlations in the disorder potential.
The inverse relative strength of the disorder is measured by the ratio
$t/W$. We consider first the extreme limit of zero hopping,
$t/W \rightarrow 0$. In that case,
the LDOS is the on-site operator
\[
	\nu_i(\varepsilon,V_i) 
	=
	\frac{\eta/\pi}{\left(\varepsilon - V_i\right)^2 + \eta^2},
\]
where $\eta$ denotes an energy-smearing parameter. 
Physically, smearing is determined by inelastic scattering,
open sample boundary conditions, or due to the finite 
energy resolution of the probing instrument. 

At the ``band center'' $\varepsilon = 0$, the distribution function 
for disorder-averaged LDOS moments evaluates to
\begin{align}\label{LDOS_Dist_AI}
	p(\nu)
	\equiv&
	\int_{-W/2}^{W/2}
	\frac{d V}{W}
	\delta\left[\nu - \nu\left(\varepsilon,V\right) \right]
	\nonumber\\
	=&
	\frac{1}{\pi \nu^2 W}
	\sqrt{
	\frac{\nu}{\numax - \nu}
	}.
\end{align}
In this equation, the LDOS is constrained 
to the interval $\numin \leq \nu \leq \numax$, where
\begin{align}\label{HardCutoffs}
	\numin 
	= 
	\frac{4 \eta}{\pi \left(W^2 + 4 \eta^2\right)},
	\quad
	\numax 
	=
	\frac{1}{\pi \eta}.
\end{align}
Using Eq.~(\ref{LDOS_Dist_AI}), one can compute the 
disorder-averaged
moments of the LDOS,
\begin{align}\label{LDOS_Mom_SS}
	\overline{\nu^q}
	=
	\frac{\Gamma\left(q - \frac{1}{2}\right)}{W \pi^{q - 1/2}  \Gamma(q)}
	\eta^{1 - q}.
\end{align}
The average LDOS is $\overline{\nu} = 1/W$; all higher moments are proportional
to $\eta^{1 - q}$, and thus \emph{diverge} in the limit of zero energy smearing
$\eta \rightarrow 0$. This not surprising, because the energy spectrum in our
trivial toy model is discrete, so that the LDOS operator becomes a delta function
with ill-defined moments as $\eta \rightarrow 0$. The moments are dominated by
the power-law (Pareto) tail of the distribution, accumulating at the upper limit
$\nu = \numax$. By contrast, the \emph{typical} LDOS, defined as 
$\typ{\nu} = \exp(\overline{\log \nu})$ is dominated by the infrared 
\[
	\typ{\nu} = \frac{4 \eta e^2}{\pi W^2}.
\]
This vanishes in the limit $\eta \rightarrow 0$. 

We see that observables exhibit broad statistics in the single site
model, governed by the $p(\nu) \sim \nu^{-3/2}$ power-law distribution
in Eq.~(\ref{LDOS_Dist_AI}). The moments are rendered finite only by the
non-zero energy smearing parameter $\eta$. This should be compared to the 
LDOS statistics in a system with extended states and weak multifractality,
e.g.\ that characterized by the quadratic $\tau(q)$ spectrum in Eq.~(\ref{tauQuad}),
with $0 < \theta \ll 1$. It is known\cite{Boris} that the corresponding
LDOS distribution has a Gaussian bulk, with a small amplitude log-normal tail
responsible for the weak multifractality. For the metallic system,
the result is independent of energy smearing, provided that the 
thermodynamic limit is taken before the smearing is set to zero.
Returning to the toy insulator model, we observe that the \emph{global} density $\nu_G$
of states (GDOS) is self-averaging in the same limit. The GDOS
is defined via
\[
	\nu_G 
	\equiv \frac{1}{N} \sum_{i = 1}^{N} \nu_i(V_i),
\]
where $N$ denotes the number of sites. In the large $N$-limit,
the cumulant expansion can be evaluated via the saddle-point.
The cumulants of the GDOS then take the form
\[
	\overline{\left[\nu_G\right]^q_c}
	=
	N^{1 - q}\,
	\left(
	\overline{\nu^q}
	+ \ldots
	\right),
\]
where $\left[\cdots\right]^q_c$ denotes the $q^{\text{th}}$ cumulant,
and the omitted terms are smaller by positive powers of $\eta$. 
Taking the infinite system size limit $N \rightarrow \infty$ before 
sending the energy smearing to zero $\eta \rightarrow 0$ leads to the vanishing
of all $q > 1$ GDOS cumulants.

The calculations above can be extended to non-zero hopping via a locator
expansion in small $t/W$, as performed by Anderson in his original 1958 paper.\cite{Anderson58} 
This expansion can be formally summed to all orders in 1D and on the Bethe lattice,\cite{BetheLat}
but an explicit solution for the LDOS statistics is difficult to obtain this way;
see Ref.~\onlinecite{SUSY} for an alternative approach.

Altshuler and Prigodin\cite{AltPrig87} succeeded in deriving the distribution generating
disorder-averaged LDOS moments in a 1D system, which is exponentially localized
for arbitrarily weak disorder.\cite{LeeRamakrishnan}
In the thermodynamic limit for a closed sample, they obtain the ``inverse Gaussian'' distribution
\begin{align}\label{AltPrigLDOSDist}
	p(\tnu)
	=
	\sqrt{\frac{4 \eta }{\pi \epsilon}}
	\frac{1}{\tnu^{3/2}}
	\exp\left[
	-\frac{4 \eta}{\epsilon}
	\frac{(\tnu - 1)^2}{\tnu}
	\right],
\end{align}
where $\tnu \equiv \nu/\overline{\nu}$, and $\epsilon$ is the typical energy
level spacing in a localization volume; $\epsilon^{-1}$ is also the elastic
scattering lifetime.\cite{AltPrig87} 
In the limit of small smearing $\eta \ll \epsilon$, this distribution has moments
\begin{align}\label{LDOS_Mom_AltPrig}
	\overline{\tnu^q}
	&=
	\frac{4^{1 - q} \Gamma\left(q - \frac{1}{2}\right)}{\sqrt{\pi}} \left(\frac{\eta}{\epsilon}\right)^{1 - q}.
\end{align}

The exact result in Eq.~(\ref{LDOS_Mom_AltPrig}) for the 1D Anderson insulator 
exhibits the same singular dependence on the energy smearing $\eta$ as the single 
site model moment in Eq.~(\ref{LDOS_Mom_SS}).
In fact, the distributions in Eqs.~(\ref{LDOS_Dist_AI}) and (\ref{AltPrigLDOSDist})
are very similar: both feature a $\nu^{-3/2}$ power law at intermediate $\nu$,
while the exponential factor in Eq.~(\ref{AltPrigLDOSDist}) plays the role of the
hard cutoffs $\nu_{\msf{min},\msf{max}}$ in Eqs.~(\ref{LDOS_Dist_AI}) and (\ref{HardCutoffs}).
The close resemblance of the exact 1D and single site model results can be attributed
to the discrete spectrum of energy levels contributing to the LDOS in an Anderson insulator, 
with an energy level spacing determined by the localization volume $\loc^{d}$
in $d$ spatial dimensions. 

The take away is that the LDOS distribution in an Anderson insulator becomes
very broad, with a power-law tail yielding divergent moments, in the limit of  
vanishingly small energy smearing. In an STM experiment performed at ultra-low temperature,
on a large, isolated Anderson localized surface, the collected LDOS statistics
should be very sensitive to the smearing induced by the energy resolution of the 
measurement itself. 

The locally discrete energy spectrum of the LDOS in the Anderson insulator
invalidates the use of Eq.~(\ref{IPR}) as a tool to compute the multifractal
$\tau(q)$ spectrum. 
As advocated above, the shape of the LDOS distribution 
function and its sensitivity to smearing can best reveal the insulating phase.
If one insists upon computing moments, one must employ\cite{Wegner} 
\begin{align}\label{tautypIns}
	\tau^{(\mathsf{loc})}(q) 
	&\equiv 
	-\frac{d}{d \ln L}
	\,
	\overline{
	\ln\left[
	\frac{1}{\overline{\nu}}
	\int_{L^d}
	d^d\vex{r}\,
	\sum_{i}
	|\psi_i(\vex{r})|^{2 q}
	\delta\left(\varepsilon - \varepsilon_{i}\right)
	\right]
	}.
\end{align} 
Since the levels are discrete, 
\begin{align}
	\lim_{\eta \rightarrow 0}
	(\pi \eta)^{q - 1}
	\nu^q(\varepsilon,\vex{r}) 
	=
	\sum_{i}
	|\psi_{i}(\vex{r})|^{2 q}
	\delta\left(\varepsilon - \varepsilon_{i}\right).
\end{align}
Thus,
\begin{align}
	\label{tautypInsLDOS}
	\tau^{(\mathsf{loc})}(q) 
	=&
	-\frac{d}{d \ln L}
	\,
	\ln\left\{
	\int_{L^d}
	d^d\vex{r}\,
	\left[
	\lim_{\eta \rightarrow 0}
	(\pi \eta)^{q - 1}
	\overline{
	\nu^q(\varepsilon,\vex{r})
	}
	\right]
	\right\}
	\nonumber\\
	&+
	\frac{d}{d \ln L}
	\,
	\ln\left\{
	\int_{L^d}
	d^d\vex{r}\,
	\overline{
	\nu(\varepsilon,\vex{r})
	}
	\right\}.
\end{align} 
In this equation, we replace averages-of-the-logs 
with logs-of-the-average, a procedure that is legitimate 
here because spatial and disorder-averaging are expected to 
yield the same results on the insulating side. Noting that the 
LDOS moments are $L$-independent in the insulator for $L \gg \loc$, 
we obtain the expected result\cite{MFCReview} for localized states
\begin{align}
	\tau^{(\mathsf{loc})}(q) = 0,
\end{align}
computed in a well-defined $\eta \rightarrow 0$ limit.

\section{Acknowledgments}

The author thanks Kostya Kechedzhi for a collaboration that lead to this
work, and thanks Andreas Ludwig, Igor Aleiner, Pedram Roushan, Haim Beidenkopf, Emil 
Yuzbashyan, and Deepak Iyer for useful discussions. The author acknowledges support by the 
National Science Foundation under Grant No.~DMR-0547769, and by the David 
and Lucile Packard Foundation.

\appendix

\section{Discrete symmetries, random matrix classification, and disorder \label{App--Sym}}

The 10 symmetry classes of disordered Hamiltonians
(Hermitian random matrices) can be efficiently distinguished
by the presence or absence of time-reversal $\mathcal{T}$, 
particle-hole $\mathcal{P}$, and chiral/``sublattice'' symmetry
$\mathcal{C}$.\cite{Classes,BernardLeClair,SRAL08}
For the two-component Dirac Hamiltonian in Eq.~(\ref{H}), 
the definitions of these symmetries are essentially unique.
In terms of the two-component Dirac spinor $\psi$, these
appear as 
\bsub
\begin{align}
	\mathcal{T}: &&\psi &\rightarrow - i \sigh_2 \psi,\quad i \rightarrow -i
	\label{T}\\ 
	\mathcal{P}: &&\psi &\rightarrow \sigh_1 \left[\psi^\dagger\right]^\T,
	\label{P}\\ 
	\mathcal{C}: &&\psi &\rightarrow \sigh_3 \left[\psi^\dagger\right]^\T,\quad i \rightarrow -i
	\label{C}.
\end{align}
\esub
In the second quantized language, $\mathcal{T}$ and $\mathcal{C}$ are antiunitary
transformations; the unitary $\mathcal{P}$ can be taken as the product of these. 

The imposition of any one of the discrete symmetries upon the Hamiltonian 
in Eq.~(\ref{H}) in every disorder realization restricts its form, and selects a 
particular random matrix symmetry class.\cite{Ludwig94,Classes,BernardLeClair,SRAL08}
(1) $\mathcal{T}$-invariance: $\vex{A} = M = 0$, only potential disorder $\Delta_{V} \geq 0$ is allowed.
Since $\mathcal{T}^2 = -1$, this is the symplectic (spin-orbit) class AII, which is 
also the symmetry class of the (presumed $\mathcal{T}$-invariant) topological $\mathbb{Z}_2$ bulk.
(2) $\mathcal{P}$-invariance: $V = \vex{A} = 0$, only random mass disorder $\Delta_M \geq 0$ is allowed. 
Because $\mathcal{P}^2 = + 1$, this is the broken time-reversal class D. 
(3) $\mathcal{C}$-invariance: $V = M = 0$, only random vector potential disorder $\Delta_{A} \geq 0$ is allowed.
This is the broken time-reversal class AIII. (Technically, it is the ``topological''/WZW 
class $\text{AIII}_1$ in the language of Refs.~\onlinecite{SRAL08,BernardLeClair}.)

Class AII is generically realized whenever time-reversal is unbroken. 
Magnetic impurities randomly polarized parallel (perpendicular) to the TI surface
manifest as point exchange sources in the vector (mass) potentials of Eq.~(\ref{H});
we are thus tempted to identify symmetry classes D and AIII with these two limits.
However, a magnetic impurity will typically induce a local potential fluctuation $V(\vex{r})$
as well. As a consequence, the generic case of broken-time reversal symmetry corresponds
to the absence of $\mathcal{T},\mathcal{P}$, and $\mathcal{C}$, which gives the unitary class A.\cite{Classes,BernardLeClair,SRAL08}
In fact, for a vanishing average mass $\overline{M} = 0$, the surface of a topological
insulator with generic time-reversal breaking disorder is expected to flow under
renormalization to the critical point of the integer quantum Hall plateau transition.\cite{Ludwig94,NomuraRyu08}
On a different note, the class $\text{AIII}_1$ and class D versions of $H$ in Eq.~(\ref{H}) can be realized 
on the surface of a bulk $\mathcal{T}$-invariant 3D topological superconductor, where time-reversal 
is respectively preserved or broken at the surface.\cite{SRAL08}

\section{Perturbation theory\label{App--PT}}

\subsection{Chiral Decomposition and one-loop renormalization}

We write a 2+0-D fermion path integral to represent correlation functions
in the disordered Dirac Hamiltonian transcribed in Eq.~(\ref{H}). The
fermion operators are replaced with the Grassmann fields 
$\{\psi,\psi^\dagger\} \rightarrow \{\psi_i,\bar{\psi}_i\}$;
here $i \in \{1,\ldots,n\}$ denotes a replica index, and we
are to send $n \rightarrow 0$ at the end of the calculation.\cite{Wegner,LeeRamakrishnan}
We employ a ``chiral decomposition'' of the two-component spinors,
\begin{align}\label{ChiralDecomp}
	\psi_i = 
	\begin{bmatrix}
	L_i \\
	R_i
	\end{bmatrix},
	\quad
	\bar{\psi}_i
	=
	\begin{bmatrix}
	\bar{R}_i
	&
	\bar{L}_i
	\end{bmatrix}
\end{align}
Then the action of the replicated
theory is 
\begin{align}\label{S}
	\mathcal{S}
	=
	\int 
	\!
	d^2\vex{r}
	\!
	\left[
	\begin{aligned}
	&
	\varepsilon
	\left(
	\bar{R}_i L_i
	+
	\bar{L}_i R_i
	\right)
	+
	\bar{R}_i L_i \phi
	+
	\bar{L}_i R_i \bar{\phi}
	\\&
	+
	\bar{R}_i
	\left(
	-i \partial
	+ A
	\right)
	R_i
	+
	\bar{L}_i
	\left(
	-i \bar{\partial}
	+ \bar{A}
	\right)
	L_i
	\end{aligned}
	\right],
\end{align}
where we have introduced complex coordinates
$\{z,\bar{z}\} = x \pm i y$, 
$\{\partial,\bar{\partial}\} = (\partial_x \mp i \partial_y)$,
and disorder potentials
$\{A,\bar{A}\} = A_x \mp i A_y$, 
$\{\phi,\bar{\phi}\} = V \pm M$.
The energy $\varepsilon$ is a fixed parameter. 
In Eq.~(\ref{S}), repeated replica
indices are summed.
Assuming the Gaussian white-noise variances
for the disorder potentials enumerated in 
Eq.~(\ref{WhiteNoise}), the replicated theory
can be averaged over disorder configurations.
The post-ensemble averaged action is 
\begin{align}\label{SBarDef}
\overline{\mathcal{S}}
	=
	\mathcal{S}_0
	+
	\overline{\mathcal{S}}_A
	+
	\overline{\mathcal{S}}_1
	+
	\overline{\mathcal{S}}_2,
\end{align}
where $\mathcal{S}_0$ is the clean Dirac action, and 
\bsub\label{SBarComp}
\begin{align}
	\overline{\mathcal{S}}_A
	=&
	-
	2 \Delta_A
	\int 
	d^2\vex{r}\,
	\bar{R}_i R_i
	\bar{L}_j L_j,
	\label{SA}
	\\
	\overline{\mathcal{S}}_1
	=&
	-
	\frac{\Delta_{V} + \Delta_M}{2}
	\int 
	d^2\vex{r}
	\,
	\left(
	\bar{R}_i L_i
	\bar{R}_j L_j
	+
	\bar{L}_i R_i
	\bar{L}_j R_j
	\right),
	\label{S1}
	\\
	\overline{\mathcal{S}}_2
	=&
	-
	(\Delta_{V} - \Delta_M)
	\int 
	d^2\vex{r}
	\,
	\bar{R}_i L_i
	\bar{L}_j R_j.
	\label{S2}
\end{align}
\esub
Different replicas become coupled through the disorder.\cite{Wegner,LeeRamakrishnan}

The disorder-averaged composite LDOS $\overline{\nu}(\varepsilon,\vex{r})$ corresponds to the fermion bilinear
expectation
\begin{align}\label{LDOSChi} 
	\overline{\nu} = \langle \bar{\psi} \psi \rangle= \langle \bar{R} L + \bar{L} R \rangle.
\end{align}
The spin LDOS $\nu^{\hat{\iota}}(\varepsilon,\vex{r})$ 
was defined by Eq.~(\ref{SpinLDOS}). For the out-of-plane 
and in-plane (chiral) components, one has 
\bsub\label{SpinLDOSChi}
\begin{align}
	\overline{\nu^{\hat{3}}} =& \langle \bar{\psi} \sigh_3 \psi \rangle = \langle \bar{R} L - \bar{L} R\rangle,\label{SpinLDOSChi3}
	\\
	\overline{\nu^{\pm}} \equiv& 
	\langle \bar{\psi} \sigh_{\pm} \psi \rangle = 2 \{\langle \bar{R} R \rangle, \langle \bar{L} L \rangle\},
	\label{SpinLDOSChipm}
\end{align}
\esub
The overlines appearing in the left-hand sides of Eqs.~(\ref{LDOSChi}) and 
(\ref{SpinLDOSChi}) denote disorder-averaging, whereas the angle brackets
on the right-hand sides represent integration in the fermion path integral, 
using the action $\bar{S}$ in Eq.~(\ref{SBarDef}).

A generic local operator corresponding to the $q^{\text{th}}$ moment of 
some fermion bilinear can be viewed as sum of ``strings,''
where each string consists of $2 q$ total right (R) and left (L) mover labels, arranged in some order.
For example, the disorder-averaged $q^{\text{th}}$ moment of the LDOS is represented by the 
the composite operator expectation value
\begin{align}\label{LDOSMomentChiral}
	\overline{\nu^q(\vex{r})}
	=
	\left\langle
	\prod_{i = 1}^{q}
	\left[
	\bar{R}_i L_i(\vex{r})
	+
	\bar{L}_i R_i(\vex{r})
	\right]
	\right\rangle.
\end{align}	
In this equation, a product is taken over operators carrying
indices in the first $q \leq n$ replicas. The $q^{\text{th}}$ 
LDOS moment is computed by placing one copy of the LDOS operator
into each of $q$ different replicas; before averaging,
this gives $\nu^q$ in a fixed realization of the disorder. 
(Placing instead the $q$ copies into the same replica would give 
the disorder-averaged first moment of a $2q$-point Green's function.) 
The operator in Eq.~(\ref{LDOSMomentChiral}) 
is an even weight sum of $2^q$ ``strings'', all of length $2 q$. 
I.e., 
\begin{align}
	\overline{\nu^q(\vex{r})}
	=&
	\{\bar{R} L ; \bar{R} L; \ldots ; \bar{R} L\}
	\nonumber\\
	&+
	\{\bar{L} R ; \bar{R} L; \bar{R} L; \ldots ; \bar{R} L\}
	\nonumber\\
	&+
	\{\bar{R} L ; \bar{L} R; \bar{R} L; \ldots ; \bar{R} L\}
	\nonumber\\
	&+
	\ldots
	\nonumber\\
	&+
	\{\bar{L} R; \bar{L} R; \ldots ; \bar{L} R\}.
\end{align}
The semicolons separate fermion bilinears in different replicas. 
Each bilinear has two entries, corresponding to the chiral identity
of the barred and unbarred operators. 

The set of all length $2 q$ strings forms a complete basis for $q^{\text{th}}$ 
moment local operators (without derivatives). These basis strings mix under renormalization
due to the disorder.\cite{Footnote--Rotations}
In general, the composite operator ($\equiv$ weighted string sum) corresponding
to the $q^{\text{th}}$ moment of a bilinear as in Eq.~(\ref{LDOSMomentChiral})
does not constitute an eigenoperator of the renormalization group.
The main task is to (1) identify 
RG eigenoperators for each disorder type and compute the spectrum of 
scaling dimensions, and (2) compute the projection
of the LDOS and (for broken $\mathcal{T}$) spin LDOS moment operators onto 
this eigenbasis, and determine the most relevant contributions.

\subsubsection{Effective Hamiltonian for 1-loop renormalization}

It is useful to view each string as a configuration of $2 q$ spin 1/2 moments. 
We associate $\{\bar{R},R\} \rightarrow 1/2$ (spin up) and 
$\{\bar{L},L\} \rightarrow -1/2$ (spin down). Operators invariant under spatial
rotations have equal numbers of up and down spins, and therefore reside in the zero total 
magnetization sector with $J^{z} = 0$. We picture each string as a basis state  
for a length $q$ chain, with two spins per site. Sites are labeled by the 
replica index $i \in \{1,\ldots,q\}$. The two spins at each site are 
distinguished by labels ``A'' and ``B,'' corresponding to barred and unbarred
operators, respectively.

Renormalization occurs via the action of the disorder vertices
appearing in Eq.~(\ref{SBarComp}), employing the clean Dirac propagator 
in a standard loop expansion. Operator mixing at one-loop is encoded in 
the effective ``Hamiltonian''
\begin{align}\label{Heff}
	H^{(\msf{eff})}
	=&
	\frac{\ln \Lambda}{2 \pi}
	\!
	\left[
	\begin{aligned}
	&
	2 \Delta_A 
	\sum_{i,j=1}^q
	\left(	
	S_{A i}^{z}
	-
	S_{B i}^{z} 
	\right)
	\left(	
	S_{A j}^{z}
	-
	S_{B j}^{z} 
	\right)
	\\
	&
	+
	\left(\Delta_M + \Delta_V\right)
	\sum_{i,j=1}^q
	\left(	
	S_{A i}^{+} S_{B j}^{-}
	+
	S_{B i}^{+} S_{A j}^{-}
	\right)
	\\
	&
	+
	\left(\Delta_M - \Delta_V\right)
	\sum_{
	\stackrel[i \neq j]{}{i,j=1}
	}^q
	\left(	
	S_{A i}^{+} S_{A j}^{-}
	+
	S_{B i}^{+} S_{B j}^{-}
	\right)
	\end{aligned}
	\right]\!\!.
\end{align}	
In this equation, $S_{A/B i}^a$ denotes a spin-1/2 operator 
acting on the barred ($A$) or unbarred ($B$) spin in replica $i$.
The prefactor obtains from evaluating the loop integrals
using a hard momentum cutoff $\Lambda$. 
The first, second, and third lines in the heavy brackets
arise through the action of the disorder vertices in $\overline{S}_A$,
$\overline{\mathcal{S}}_1$, and $\overline{\mathcal{S}}_2$, respectively. 
The $\Delta_A$ renormalization is diagonal in the $\uparrow$/$\downarrow$
($R/L$) basis. By contrast, $\overline{\mathcal{S}}_1$, and $\overline{\mathcal{S}}_2$
perform single exchanges of right and left labels.
$\overline{\mathcal{S}}_1$ ($\overline{\mathcal{S}}_2$) mediates interflavor $A\leftrightarrow B$
(intraflavor $A\leftrightarrow A$, $B\leftrightarrow B$) exchanges. 
Summing the angular momenta,
\begin{align}\label{HeffSummed}
	H^{(\msf{eff})}
	=&
	\frac{\ln \Lambda}{2 \pi}
	\!
	\left\{
	\begin{aligned}
	&
	2 \Delta_A 
	\left(	
	J_{A}^{z}
	-
	J_{B}^{z} 
	\right)^2
	\\
	&
	+
	2\left(\Delta_M + \Delta_V\right)
	\left(	
	J_{A}^{x} J_{B}^{x}
	+
	J_{A}^{y} J_{B}^{y}
	\right)
	\\
	&
	+
	\left(\Delta_M - \Delta_V\right)
	\\
	&
	\phantom{+}
	\times
	\left[
	\begin{aligned}
	&
	\vex{J}_A^2 - \left(J_{A}^z\right)^2
	+
	\vex{J}_B^2 - \left(J_{B}^z\right)^2
	\\&
	- 
	q
	\end{aligned}
	\right]
	\end{aligned}
	\right\},
\end{align}
where 
$\vex{J}_{A,B} \equiv \sum_{i} \vex{S}_{A,B i}$.

In the general case of broken $\mathcal{T}$ discussed
in 
Sec.~\ref{ClassA}, 
all three disorder parameters are present. 
The most relevant eigenvalue of Eq.~(\ref{HeffSummed}) determines
the scaling of the $q^{\text{th}}$ LDOS moment.\cite{Footnote--Parity}
The first few multifractal moments for various disorder configurations were
obtained through numerical diagonalization; results appear in Figs.~\ref{FigRenyi2} 
and \ref{FigRenyi3}. 

Even moments of the out-of-plane spin LDOS $\overline{\nu^{\hat{3}}}$ 
[Eq.~(\ref{SpinLDOSChi3})] are invariant under spatial rotations and 
parity.\cite{Footnote--Parity} 
In the multidisorder unitary case, 
even moments of the composite $\nu$ and out-of-plane spin $\nu^{\hat{3}}$ LDOS
are dominated by the same RG eigenoperator.
The dimension $x_q^{\hat{3}}$ that determines the spin LDOS scaling
via Eq.~(\ref{ISPR--Scaling}) is the same that enters into 
the LDOS spectrum in Eq.~(\ref{tauScalingDim}), Figs.~\ref{FigRenyi2} and \ref{FigRenyi3}.
Moments of the chiral spin LDOS components in Eq.~(\ref{SpinLDOSChipm}) correspond to the
highest weight states $\ket{j = q; m = \pm q}$; here, $j(j+1)$ denotes the eigenvalue
of $(\vex{J}_A + \vex{J}_B)^2$, with $0 \leq j \leq q$ and $-j \leq m \leq j$. 
These highest weight states are annihilated by $H^{(\msf{eff})}$ in Eq.~(\ref{HeffSummed}), 
leading to Eq.~(\ref{x_q^pm A}).

Below we discuss the special cases of isolated disorder flavors.


\subsubsection{Broken $\mathcal{T}$: random vector potential disorder (Class AIII)}

For $\Delta_M = \Delta_V = 0$, Eq.~(\ref{HeffSummed}) reduces to
\begin{align}\label{HeffA}
	H_A^{(\msf{eff})}
	=&
	\frac{\ln \Lambda}{\pi}
	\Delta_A 
	\left(	
	J_{A}^{z}
	-
	J_{B}^{z} 
	\right)^2
	\nonumber\\
	=&
	\frac{\ln \Lambda}{\pi}
	\Delta_A 
	\left(m_A - m_B\right)^2.
\end{align}
On the second line, we have evaluated $H_A^{(\msf{eff})}$
for the product state $\ket{j_A, j_B; m_A, m_B}$.
Since $\max(j_{A,B}) = q/2$ and $|m_{A/B}| \leq j_{A/B}$,
the maximum eigenvalue attains for the states 
$\ket{q/2, q/2 ; q/2, -q/2} \rightarrow \{\bar{R} L; \bar{R} L; \ldots ; \bar{R} L\}$ 
and $\ket{q/2, q/2 ; -q/2, q/2} \rightarrow \{\bar{L} R; \bar{L} R; \ldots; \bar{L} R\}$.
These have $J^z = 0$, and thus correspond to operators invariant under spatial rotations;
the symmetric combination is also parity-invariant.\cite{Footnote--Parity}
Via standard renormalization group machinery,\cite{Amit,Foster08}
one obtains the most relevant scaling dimension for a $q-$fold product
operator,
\begin{align}\label{ScalingDimVector}
	x^{(A)}_{q}
	=&
	q
	- 
	q^2
	\frac{\Delta_A}{\pi}.
\end{align}
Using Eq.~(\ref{ScalingDimVector}) in Eq.~(\ref{tauScalingDim}) gives the
result for the quadratic $\tau(q)$ spectrum in Eqs.~(\ref{tauQuad}) and (\ref{tauAIII}).

\subsubsection{Broken $\mathcal{T}$: random mass disorder (Class D)}

For the random mass case, Eq.~(\ref{HeffSummed}) becomes
\begin{align}\label{HmFinal}
	H_{M}^{(\msf{eff})}  
	=&
	\frac{\Delta_M \ln \Lambda}{2 \pi}
	\left[
	\vex{J}^2
	-
	(J^{z})^2
	- 
	q
	\right]
	\nonumber\\
	=&
	\frac{\Delta_M \ln \Lambda}{2 \pi}
	\left[
	j(j+1)
	-
	m^2
	- 
	q
	\right],
\end{align}
where $\vex{J} \equiv \vex{J}_A + \vex{J}_B$.
On the second line, we have evaluated the ``Hamiltonian'' on a total angular momentum
eigenstate $\ket{j m}$. For $2 q$ spins, we have $\max(j) = q$.
The maximum eigenvalue is associated to the non-degenerate $j = q$, $m = 0$ state,
which is invariant under spatial rotations. The scaling dimension is
\begin{align}\label{ScalingDimMass}
	x^{(M)}_{q}
	=&
	q
	- 
	q^2
	\frac{\Delta_M}{2 \pi}.
\end{align}
The corresponding eigenoperator $\ket{j = q, m = 0}$ is an equal weight symmetric sum of all
permutations of $q$ ``$R$'' and $q$ ``$L$'' labels, and has non-zero overlap with the 
naive LDOS moment (a $q$-fold triplet product) in Eq.~(\ref{LDOSMomentChiral}).
Using Eq.~(\ref{ScalingDimMass}) in Eq.~(\ref{tauScalingDim}), one obtains the
result for the quadratic $\tau(q)$ LDOS spectrum in Eqs.~(\ref{tauQuad}) and (\ref{tauD}).
By contrast, the out-of-plane spin LDOS (mass operator) in Eq.~(\ref{SpinLDOSChi3})
is a singlet; the disorder-averaged $q^{\text{th}}$ moment thus corresponds to the 
eigenoperator $\ket{j = 0, m = 0}$ [leading to Eq.~(\ref{x_q^3 D})].

\subsubsection{Non-magnetic disorder (Class AII)}

We rotate the ``B'' composite spin by $\pi$ around the $\hat{z}$-axis, 
\begin{align}
	J_B^{x}
	\rightarrow
	\tilde{J}_B^{x}
	\equiv
	- J_B^{x},
	\quad
	J_B^{y}
	\rightarrow
	\tilde{J}_B^y
	\equiv
	- J_B^{y},
	\quad
	J_B^{z}
	\rightarrow
	\tilde{J}_B^{z}
	\equiv
	J_B^z.
	\nonumber
\end{align}
The Hamiltonian in Eq.~(\ref{HeffSummed}) with $\Delta_M = \Delta_A = 0$ 
becomes
\begin{align}\label{Hv-2}
	H_{V}^{(\msf{eff})}
	=&
	-
	\frac{\Delta_{V} \ln \Lambda}{2 \pi}
	\left[
	\tilde{\vex{J}}^2
	- 
	(\tilde{J}^{z})^2 
	-
	q
	\right],
\end{align}
where $\tilde{\vex{J}} \equiv \vex{J}_A + \tilde{\vex{J}}_B$.
Eq.~(\ref{Hv-2}) has the same form as Eq.~(\ref{HmFinal}), with 
$\Delta_{V} \rightarrow - \Delta_M$.
This is consistent with a mapping between the random mass and
vector potential models identified in Ref.~\onlinecite{Ludwig94}.
In the case of the scalar potential, the maximum eigenvalue is associated
to the highly degenerate singlet sector $\ket{\tilde{j} = 0, \tilde{m} = 0}$,
leading to the scaling dimension
\begin{align}
	x^{(V)}_{q}
	=&
	q
	- 
	q
	\frac{\Delta_{V}}{2 \pi}.
\end{align}
Using this result in Eq.~(\ref{tauScalingDim})
gives $\tau(q) = 2(q - 1)$. We conclude that no moment operator (without derivatives)
acquires multifractal scaling at one loop for the $\mathcal{T}$-invariant
model.

\subsection{Two loop renormalization, $\mathcal{T}$-invariant class AII}

In the $\mathcal{T}$-invariant class AII model, the first correction to the LDOS $\tau(q)$ 
spectrum appears at second order in $\Delta_{V}$. We have carried out a two-loop calculation
and found that the naive LDOS moment in Eq.~(\ref{LDOSMomentChiral}) remains an eigenoperator.
To this order, we find the scaling dimension
\begin{align}\label{xv2loop}
	x^{(V)}_{q}
	=&
	q
	- 
	q
	\frac{\Delta_{V}}{2 \pi}
	-
	\frac{\Delta_{V}^2}{8 \pi^2}
	\left[
	3 q(q-1) + q
	\right]
	+ 
	\ord{\Delta_{V}^3}.
\end{align}
Combining Eqs.~(\ref{tauScalingDim}) and (\ref{xv2loop}),
we recover the quadratic multifractality for $\tau(q)$ quoted
in the text, Eqs.~(\ref{tauQuad}) and (\ref{tauAII}).
We have used dimensional regularization to obtain the result
in Eq.~(\ref{xv2loop}). Although the Clifford algebra
becomes formally 
infinite\cite{Bondi90-BennettGracey99} 
upon dimensional continuation to $d = 2 - \epsilon$, this 
causes no problems for the renormalization of the $q^{\text{th}}$ 
LDOS moment because the latter is already an eigenoperator. 
We omit details of the (lengthy) two-loop calculation in this paper.


\begin{thebibliography}{99}
\bibitem{Footnote--IsolatedDisorder}
	As discussed 
	in 
	the paragraphs following 
	Eq.~(\ref{Delta_v Screened}) in Sec.~\ref{Model}, a magnetic 
	impurity will typically induce a scalar potential $V(\vex{r})$ 
	deformation, in addition to mass and vector potential point exchanges for
	out-of-plane and in-plane polarization components, respectively.
\bibitem{Footnote--Parity}
	For composite LDOS fluctuations, it is necessary to restrict the
	eigenspace of Eq.~(\ref{Heff}) to rotationally invariant ($J^z = 0$),
	``parity''-invariant states.
	Here, parity denotes simultaneous invariance under spatial $x$- and 
	$y$-reflections in the plane of the system.
	In the chiral decomposition of Eq.~(\ref{ChiralDecomp}),
	parity-invariant operators are symmetric under the exchange
	$R \leftrightarrow L$, $\bar{R} \leftrightarrow \bar{L}$.
\end{thebibliography}
\end{document}